\documentclass{iopjournal}

\usepackage[numbers]{natbib}

\begin{document}

\articletype{paper} %	 e.g. Paper, Letter, Topical Review...

\title{Experimental evidence of $\mathrm{T}_\mathrm{c}$ enhancement above 50 K and diode and paramagnetic-Meissner effects,
in Nickelate films on highly reduced SrTiO$_3$}

\author{Anna Eyal and Gad Koren$^*$}
%$\orcid{0000-0000-0000-0000}, Author Name$^2$\orcid{0000-0000-0000-0000} and Author Name$^{1,*}$\orcid{0000-0000-0000-0000}}

\affil{Physics Department, Technion, Haifa, 32000, Israel }

%\affil{$^2$Department, Institution, City, Country}

\affil{$^*$Author to whom any correspondence should be addressed.}

\email{gkoren@physics.technion.ac.il}

\keywords{Nickelate films, Superconductivity, Diode effect, Paramagnetic Meissner effect}

\begin{abstract}
Since the discovery of superconductivity in nickelate thin films in 2019, the quest for enhancing their $\mathrm{T}_\mathrm{c}$ has been ongoing. Here we provide experimental evidence for $\mathrm{T}_\mathrm{c}$ enhancement in oxygen deficient films on highly reduced and conducting SrTiO$_3$ substrates. $\mathrm{T}_\mathrm{c}$ onset of 50-70 K was found in Meissner and transport measurements, which indicates superconductivity in islands or domains in our films, where $\mathrm{T}_\mathrm{c}$ of zero resistance is obtained at 20-25 K. In addition, we observed  a giant paramagnetic-Meissner effect peak at about 48 K, which further supports the existence of a superconductive transition just above it. Furthermore, an asymmetric or nonreciprocal and non-hysteretic superconductive diode effect was observed. The latter effect could be fully polarized, and its polarity could be reversed. Our mixed phase films comprised of many Ruddlesden-Popper $\rm (Nd_{0.8}Sr_{0.2})_{n+1}Ni_nO_{3n+1}$ phases and includes the infinite-layer (IL) phase. 
\end{abstract}

\section{Introduction}

The discovery of superconductivity in nickelate thin films by the Hwang group in 2019 \cite{Hwang} ignited worldwide interest in this new family of materials \cite{MikeNorman,GoodgeNorman2025,LeeGoodge,LiHwangPRL,OsadaHwang,LiWen,Zhou2020,Zhou2024,Ariando,GuWen,LeeHwang}. Given their similarity to cuprates, nickelates were immediately recognized as a potential new platform for high-temperature superconductivity, as discussed in an early review by Norman \cite{MikeNorman}, and more recently based on up to date results by Goodge and Norman \cite{GoodgeNorman2025}. Initial studies focused on mapping the phase diagram of these Nd$_{0.8}$Sr$_{0.2}$NiO$_2$ (112 phase) films and characterizing their fundamental properties \cite{LiHwangPRL}. Soon after, superconductivity was also observed in Pr$_{0.8}$Sr$_{0.2}$NiO$_2$ thin films \cite{OsadaHwang}, reinforcing the idea that superconductivity could emerge in multiple members of the nickelate family. However, in both this study and in Ref. \cite{Hwang} the films thickness ranged between 5 and 12 nm only. Thicker films of this kind showed no superconductivity in bulk form \cite{LiWen,Zhou2020}, and raised the question whether the observed superconductivity in thin films of a few nanometers is an interface effect originating in strains or other interactions with the substrate. Actually, strain of the films with the substrate was found to change their $\rm T_c$ very significantly, though limits to this strain engineering effect was also demonstrated \cite{Segedin2023}.  More recently, fabrication of free-standing superconducting nickelate films was demonstrated, and a maximum $\mathrm{T}_\mathrm{c}$ of 25 K was found in films on (110) $\rm NdGaO_3$ \cite{FreeStanding}. Even higher superconducting $\mathrm{T}_\mathrm{c}$ values above 30 K at ambient pressure were obtained in ultra thin films of $\rm Sm_{1-x-y-z}Eu_xCa_ySr_zNiO_2$ also on $\rm NdGaO_3$ wafers \cite{ChowAriando}. The highest $\mathrm{T}_\mathrm{c}$ so far, of approximately 80 K, was found in well oxygenated single crystals of the double-layer nickelate La$_3$Ni$_2$O$_7$ (327 phase) under high pressure of tens of GPa \cite{HighPressure,Li327press}. Pressure turned out to be a very effective lever to control $\rm T_c$ of the nickelates, the cuprates and other superconductors.  Even in the relatively low $\rm T_c$ of the 112 IL films, it was demonstrated very recently that such free standing films under 90 GP pressure attained $\rm T_c$ onset of 74 K \cite{Lee112press}. Fig. 3 of this study that depicts $\rm T_c$ versus pressure, shows a very instructive comparison with other superconductors. So strains and pressure have similar and complementary effects on the $\rm T_c$ of superconductors.\\   

From a practical point of view, increasing $\rm T_c$ at ambient pressure is preferable. 
Four studies recently reported ambient pressure superconducting $\mathrm{T}_\mathrm{c}$ onset over 40 K in the bilayer 327 phase of nickelate films \cite{Ko,Zhou2024,Liu,Haw2025}. In these works, an oxidation annealing step in Ozone was employed in order to get the high $\mathrm{T}_\mathrm{c}$ phase, unlike previous works on nickelate films where  a reduction annealing process, generally in $\rm CaH_2$, was used. The highest $\mathrm{T}_\mathrm{c}$ onset in these studies was obtained in compressively strained La$_2$PrNi$_2$O$_7$ films with $\mathrm{T}_\mathrm{c}$ onset of 48 K by Liu et al. \cite{Liu}. A more detailed study of strain effects in these films on different substrates was recently reported by the same group \cite{Tarn2025}.  \\

In earlier studies, most films showing signs of superconductivity were prepared in a two-step process, whereby the perovskite films were deposited first, followed by a second step of reduction in $\rm CaH_2$. These films, usually with thicknesses of a few to several nm, were comprised of the infinite-layer 112 phase, possibly in close proximity to the substrate, as the carrier of the superconductive currents. In the study of films of the 327 bilayer phase \cite{Liu}, laser deposition was still used but at a much lower fluence on the target and this was followed by Ozone oxidation. In the present study, we followed the earlier two-steps route and prepared the perovskite phase first, like other groups did. But unlike other groups, the reduction in $\rm CaH_2$ step was carried out under pumping. This led to an effective reduction not only of the films but also of the $\rm SrTiO_3$ (STO) substrate, which became black and highly conductive. After this annealing process, magnetic moment measurements versus temperature showed a $\mathrm{T}_\mathrm{c}$ onset at about 50 K. Furthermore, resistance versus temperature of our films showed a $\mathrm{T}_\mathrm{c}$ onset of 60-70 K and $\mathrm{T}_\mathrm{c}$ offset with zero resistance at 20-25 K. Therefore, the higher $\mathrm{T}_\mathrm{c}$ indicates superconductivity in islands, where the lower transition to zero resistance superconductivity is due to the weak-links between them becoming superconducting as well.\\

In this study we also observed two additional effects. One is a highly asymmetric diode effect in the voltage versus current curves (VICs) without an external magnetic field, where under one polarity of the bias current an Ohmic behavior was found, while in the opposite polarity a supercurrent was observed. However, the sign of this effect could be reversed depending on the distance between the contacts used. This kind of diode effect was recently observed by Qi et al. in BSCCO \cite{Qi}, by Le et al. in a Kagome superconductor \cite{Le_Kagome} and theoretically investigated by Mori et al. where a sign reversal was found \cite{Mori}. The other observation we made is of a giant paramagnetic Meissner effect peak (PME, also known as the Wohlleben effect \cite{Wohlleben}), where a positive magnetization peak appears just below the superconducting transition at about 50 K. This might indicate odd-frequency superconductivity \cite{Linder}, the presence of SN or SF junctions in our samples \cite{Koblischka} or the development of a giant vortex state \cite{Moshchalkov}. The PME peaks were most prominent under magnetic fields of thousands Oersted normal to the wafer, unlike earlier studies where fields around one Oersted were used \cite{Wohlleben}.  \\   

We believe that the main reason for our extraordinary results originates in the combination of highly reduced black STO substrates and the nickelate films. Stand-alone wafers of pure black STO were metallic, highly conductive and had a superconducting transition at $\mathrm{T}_\mathrm{c}$ of 0.25 K, similarly to optimally doped reduced STO \cite{blackSTO}. When short reduction time was used, we identified the superconductive phase by X-ray diffraction (XRD) as due to the infinite-layer  Nd$_{0.8}$Sr$_{0.2}$NiO$_2$ as other groups have found, but when the reduction time was increased, XRD of our films showed no particular features beyond the almost disappearance of the perovskite phase. It turned out that in this case, our films became polycrystalline, thus decreasing the XRD signals to their noise level. In both type of cases though, a signature of $\mathrm{T}_\mathrm{c}$ above 50 K was observed. However, high-angle annular dark-field scanning transmission electron microscopy (HAADF-STEM) measurements revealed a more complex picture as for the structure and composition of our films. Basically, it showed that under a short annealing time, our aged reduced films comprised of a mixture of Ruddlesden-Popper (RP) phases  $\rm (Nd_{0.8}Sr_{0.2})_{n+1}Ni_nO_{3n+1}$
with a dominant 113 phase (n$\rightarrow\infty$), smaller amounts of lower n phases, and a residual IL phase. When longer annealing times were used, the films became polycrystalline with more prominent n=2, 3 and 10 phases, and without clear phases of 113 or IL.  Note that due to aging of the STEM samples and oxygen intake, very little remained of the IL phase in our film \cite{Rossi2024,Parzyck2024,Raji2023,Flavenot}.  It should also be noted here that our freshly reduced films could also be of the square planar type $\rm (Nd_{0.8}Sr_{0.2})_{n+1}Ni_nO_{2n+2}$ that when n$\rightarrow\infty$ becomes the infinite-layer 112 phase \cite{GoodgeNorman2025,Segedin2023}. \\

\section{Experimental}

All our Nickelate films were prepared by pulsed laser deposition (PLD) on optically polished Ti-terminated (100) STO wafers of 10$\times$10 mm$^2$ area, using a ceramic target of Nd$_{0.8}$Sr$_{0.2}$NiO$_3$ (NSNO). 
In our previous work on NSNO in 2021 \cite{Koren21} we used the third-harmonic of a Nd-YAG laser with 355 nm wavelength to deposit the films. This did not yield a dominant perovskite phase of the Nickelate films, as the (001) XRD peak of these films was absent or too weak to be seen. We stress that this was not due to limited resolution of our XRD measurements, since in similar manganite films of Nd$_{2/3}$Sr$_{1/3}$MnO$_3$, $all$ (00$n$) peaks were clearly resolved, including the odd-n ones. In the present study, we used the fourth-harmonic of the Nd-YAG laser with 266 nm wavelength and this yielded the desired perovskite 113 phase with all the (00$n$) peaks as depicted in Fig. 1 (a)-(c). We point out that these film peaks are shifted slightly to the right compared to the expected 113 peak positions (the vertical blue bars), indicating a smaller c-axis lattice constant, as expected due to tensile strain with the STO substrate (a=0.3905 nm) which leads to shrinking of the c-axis. Going back to the PLD process, apparently, the deeper UV wavelength is necessary for obtaining the perovskite phase in the Nickelates. This is in contrast to YBCO where all the Nd-YAG harmonics yielded a stable perovskite phase \cite{Koren89}. Films of 10, 20, 30 and 80 nm thickness were prepared, at $780\,^\circ\mathrm{C}$ heater block temperature ($\sim600\,^\circ\mathrm{C}$ wafer temperature) and 100 mTorr O$_2$ flow. The 266 nm laser operated at 3.3 Hz and was focused to 1.5-2.5 J/cm$^2$ on the target. After deposition, 10 Torr of oxygen was added and the sample was cooled to room temperature with a dwell of 1 hour at $450\,^\circ\mathrm{C}$.     \\

Reduction or annealing of the films was carried out in aluminum  foil packets with 0.35 g of $\rm CaH_2$ powder at about $320\,^\circ\mathrm{C}$, for various durations between one and six hours. Highly resistive films were obtained when using closed packets. However, when a small aperture was made in the packets and the annealing was performed under pumping in a chamber with a base pressure of 10$^{-7}$ Torr at $25\,^\circ\mathrm{C}$, low resistance films had been produced. During the actual reduction process at $320\,^\circ\mathrm{C}$, the pressure in the vacuum chamber rose to about 5-10$\times$10$^{-7}$ Torr, while inside the packet we estimate that a pressure of around 10$^{-5}$ Torr was obtained due to the differential pumping used. Unlike in other studies, under our annealing conditions, the STO wafers were also reduced and became black, metallic, and highly conductive (apparently on the surface). This was also tested on bare STO wafers without the NSNO films, and the resulting black wafers were also superconducting at 0.25 K as found in the literature for optimally reduced STO \cite{blackSTO}. The transport and magnetic results on these wafers served as reference or background for the measurements of the films themselves.   \\ 

\begin{figure}
 \centering
        \includegraphics[width=1\textwidth]{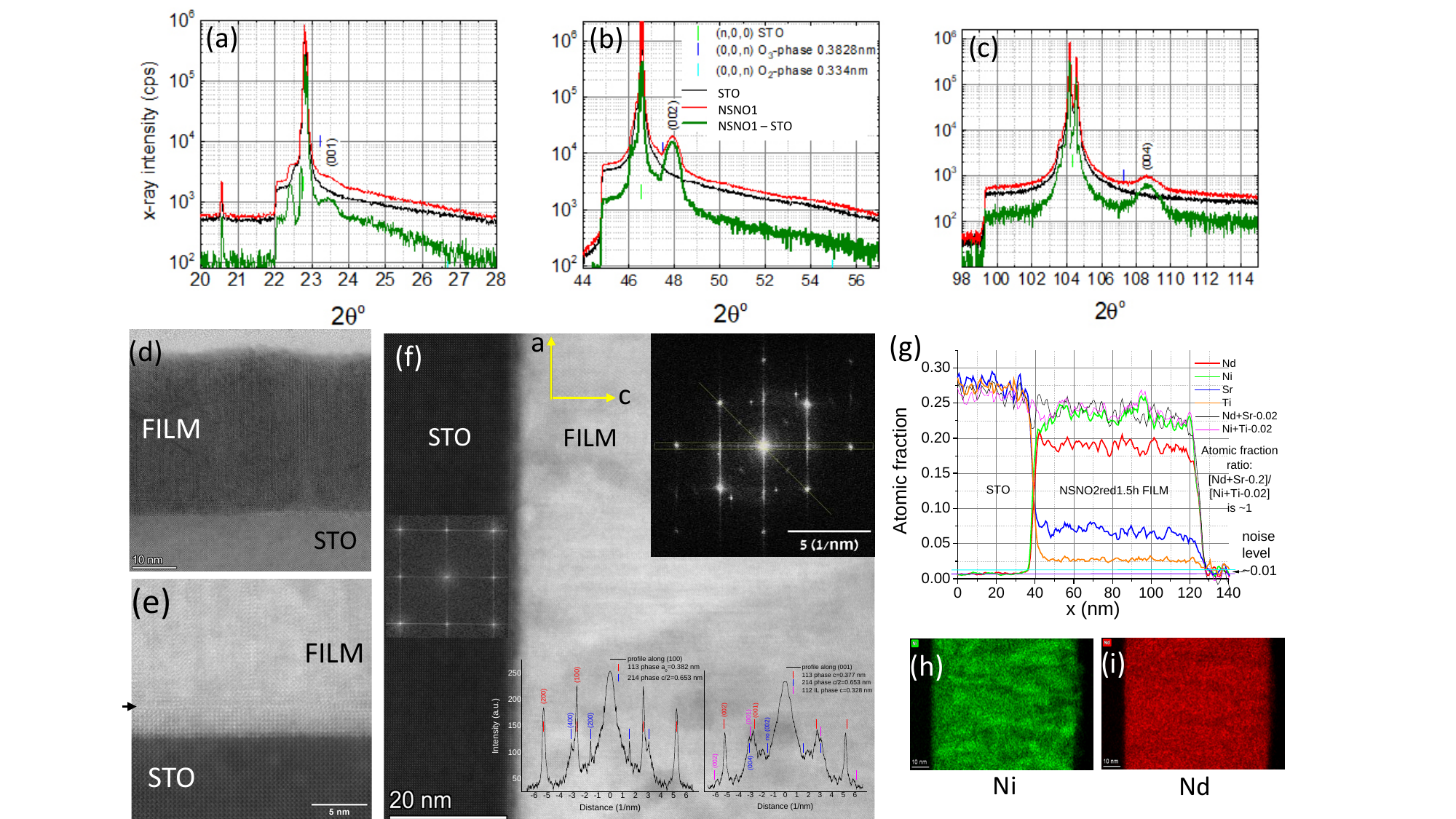}
 \caption{ XRD, HAADF-STEM and EDS results of our films. 
(a)-(c) show XRD results of the virgin NSNO1 film, where all the (00$n$) peaks of the epitaxial  $\rm Nd_{0.8}Sr_{0.2}NiO_3$ perovskite phase are seen, except for the weakest (003) peak. The green traces represent data where the STO background is subtracted. (d) and (e) show STEM images of another virgin film NSNO57 along the [100] zone axis. In (d) a  vertical columnar growth is seen while in the zoomed-in image (e) a highly ordered, 2 nm thick epitaxial perovskite layer right above the interface is observed (up to the arrow), while defects are developed further up.  (f) shows a STEM image also along the [100] zone axis, but of a reduced film NSNO2red1.5h, where the epitaxial film is now quite disordered. FFT of the whole film is shown in the top-right inset in (f). Intensity profiles along the [100] orientation (along the narrow yellow rectangle in the FFT image) and the [001] orientation are shown by the two insets at the bottom of (f). These profiles indicate the existence of 113 and 214 RP phases with possible residual 112 IL phase. Note that along the diagonal of the FFT image, the spots at $\pm$2 1/nm belong to the 327 phase. (g-i) depict EDS data of the film in (f), where in (g) the net Nd+Sr atomic fraction is equal to that of the net Ni+Ti. Ni-rich regions in the film are seen in (h), while in (i) these regions are Nd poor.  }
\label{fig1}
\end{figure}

Structural properties of our films were determined by standard $\theta$ - 2$\theta$ x-ray diffraction (XRD) measurements, and the morphology of the films by atomic force microscopy (AFM) images. We also used HAADF-STEM measurements on a Titan 300 kV microscope and a 200 kV Spectra microscope to determine the phases involved, their epitaxiality, defects and crystallinity.  Transport measurements of our films were carried out using a PPMS Dynacool system of Quantum Design. When high sensitivity was needed, the PPMS electronics was replaced by an external Keithley 2450 source, while still using the Dynacool's cryogenics. The contacts for the 4-probe measurements were made by a wire bonder with Al wire, and these were also overlayed by freshly applied silver paste bands for increasing the contacts area and thereby reducing their resistance by two orders of magnitude. Magnetic measurements were carried out using a Quantum Design MPMS3 SQUID magnetomoter. In these measurements, the samples were placed in a straw of 5 mm diameter in a way that the applied magnetic field was perpendicular to the film's surface.  Only in one case, a magnetic field parallel to the film's surface was used, in order to check if our results were affected by regions of magnetic Nd, Ni or their oxides. 
\\

Overall, 12 nickelate films were prepared and characterized in this study using PLD with the fourth-harmonic of the Nd-YAG laser. The virgin films were generally cut into four quarters of 5$\times$5 mm$^2$ area, for reduction in $\rm CaH_2$ under different annealing conditions such as temperatures, pressure and time duration. In the present study we report only on the last four films that were reduced in $\rm CaH_2$ at $320\,^\circ\mathrm{C}$ and under pumping at local pressure (in the Al packet) of about 10$^{-5}$ Torr. We note here that the AFM images were taken on virgin or pristinely reduced films immediately after their preparation, while the HAADF-STEM results of the reduced films were obtained on 3-4 months aged films, either after being kept dry in a desiccator or after extensive transport and magnetic measurements. Only the STEM results of the virgin film were obtained on a freshly prepared film.

\section{Results}

\subsection{Short reduction time film}

\begin{figure}
 \centering
        \includegraphics[width=1\textwidth]{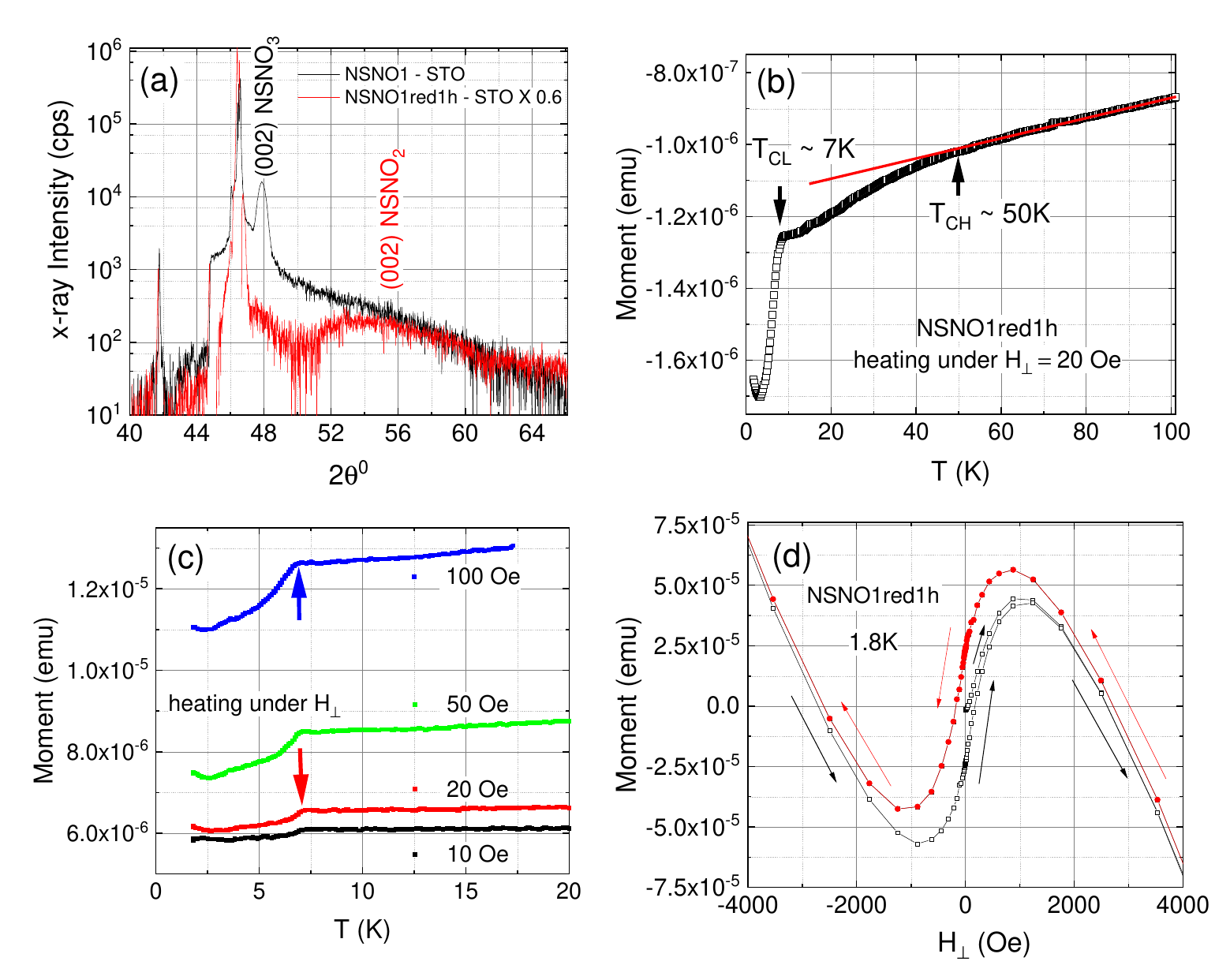}
 \caption{ XRD and magnetization results of NSNO1 and NSNO1red1h films.  XRD of the virgin NSNO1 and reduced NSNO1red1h films are shown in (a), where after annealing the prominent perovskite (002) peak of NSNO1 disappears and the broad (002) peak of the infinite-layer $\rm Nd_{0.8}Sr_{0.2}NiO_2$ phase appears. (b) and (c) show magnetic moments versus temperature of NSNO1red1h where two superconducting transitions are observed at about 7 and 50 K in (b), while in (c), scaling of the moment magnitude with field is demonstrated. (d) depicts a magnetic hysteresis loop of this film at 1.8 K after ZFC, where the field was cycled to $\pm$5000 Oe and back to 0 Oe. The same response as in (d) is obtained also on a bare STO wafer (see Fig. S2 in the supplement). }
\label{fig2}
\end{figure}

We start by presenting the structural characteristics of our films. Figure 1 (d) and (e) depict HAADF-STEM images taken along the [100] zone axis of a virgin nickelate film NSNO57 prepared under 1.5 J/cm$^2$ laser fluence. Both show epitaxial growth of this 40 nm thick film. In (d) a prominent columnar structure normal to the STO wafer is observed, while in the zoom-in image in (e) a highly ordered 2 nm thick layer of the 113 perovskite phase is clearly seen just above the interface (up to the arrow in this image). Above this layer vertical columnar defects develop as in (d). FFT analysis of this film shows that the dominant phase is the 113 RP phase where n is very large (see Fig. S6 in the supplement, and note that it is along the [110] zone axis). In this phase, a rock-salt layer of $\rm Nd_{0.8}Sr_{0.2}O$ is inserted between each n blocks of the perovskite $\rm Nd_{0.8}Sr_{0.2}NiO_3$ octahedra giving rise to the vertical columnar structure as in Fig. 1 (d). This is also consistent with STEM results on virgin films reported in the literature \cite{Hwang,LeeGoodge}.\\

In Fig. 1 (f) we present an  HAADF-STEM image along the [100] zone axis of an 80 nm thick film NSNO2red1.5h that was reduced for 1.5 h. One can see that the reduction process introduced more disordered regions in the reduced film compared to defects found in our virgin films (as in Fig. 1 (d) and (e) here and in Fig. S6 of the supplement). An FFT image of this film is shown in the top inset to this figure. Intensity profiles along the [100] and [001] directions are given by the bottom two insets. They show that the 113 RP phase is still dominant, but with other RP phases with n=1, 2 and 3 appearing, including a possible residual IL phase. The IL phase is unstable over a long time (4 months here) due to oxygen intake from the atmosphere. Nevertheless, the peak at 3 1/nm in the bottom right inset seems to be due to (001) of the infinite-layer phase and not to the overlapping the (004) peak of the 214 phase, since no (002) peak of the 214 phase is observed in this profile.\\

 Figure 1 (g)-(i) show the corresponding EDS results of the atomic fractions of the elements in the NSNO2red1.5h film and their atomic distribution maps. One can see that a small and quite uniform atomic fraction of Ti is found in this reduced film. This is obviously due to diffusion from the STO substrate, which was apparently enhanced by the reduction under vacuum used here, that removed oxygen from the STO substrate and thereby weakened the Ti bonds in it near the interface. However, a small Ti concentration was already found in our virgin (unreduced) film as depicted in Fig. S10 of the supplement. Since the atomic radii of Ni and Ti are small and comparable (0.124 and 0.147 nm, respectively), it is most likely that Ti substitutes for Ni in the NiO$_2$ planes of the films (like Ni, Co, Fe and Zn substitution in the CuO$_2$ planes of the cuprates). The atomic radii of Nd and Sr are much larger (0.182 and 0.215 nm, respectively), and these atoms occupy the space between the planes as in the infinite-layer structure \cite{GoodgeNorman2025}. Actually, the Ti concentration within the NiO$_2$ planes here in the reduced film is quite small $\sim 6 \%$, while in the virgin film of Fig. S10 it is even smaller about 3$\%$. \\

We estimate the noise level in Fig. 1 (g) as the average of the noise below the film (in the STO wafer) and above it, (in between the two horizontal lines in this figure). By subtracting this noise from the corresponding atomic fractions involved, we find that  the net average atomic fraction ratio of (Nd+Sr)/(Ni+Ti) is approximately 1. This result indicates that the target stoichiometry is preserved in the film, consistent with our previous finding that the 113 RP phase is the dominant phase in this film. However, our films are not homogeneous as seen in Fig. 1 (h), where the Ni tends to aggregate in puddles. Furthermore, this non uniformity is already present in our virgin films as seen in Fig. S9 of the supplement, where also the Nd inhomogeneity is clearly visible and alternate with that of the Ni. \\

Next we present XRD data and the magnetic properties of our 1 h reduced film.  Figure 2 (a) shows XRD results of a virgin 30 nm thick NSNO1 film together with this film after reduction in $\rm CaH_2$ at $320\,^\circ\mathrm{C}$ for 1 h to produce NSNO1red1h. One can see that the perovskite (002) peak of the virgin film disappeared in the reduced film, while a broad (002) peak of the infinite-layer $\rm Nd_{0.8}Sr_{0.2}NiO_2$ phase appeared. We stress that no other clear film's peaks were visible in the $10\,^\circ<2\theta <140\,^\circ$ range of this XRD spectrum. The magnetic moment as a function of temperature of the reduced film is shown in Fig. 2 (b), where a zero field cooling (ZFC) to 1.8 K was performed, followed by applying a magnetic field of 20 Oe and measuring the moment while heating the sample, as in standard Meissner effect measurement.\\

\begin{figure}
 \centering
        \includegraphics[width=1\textwidth]{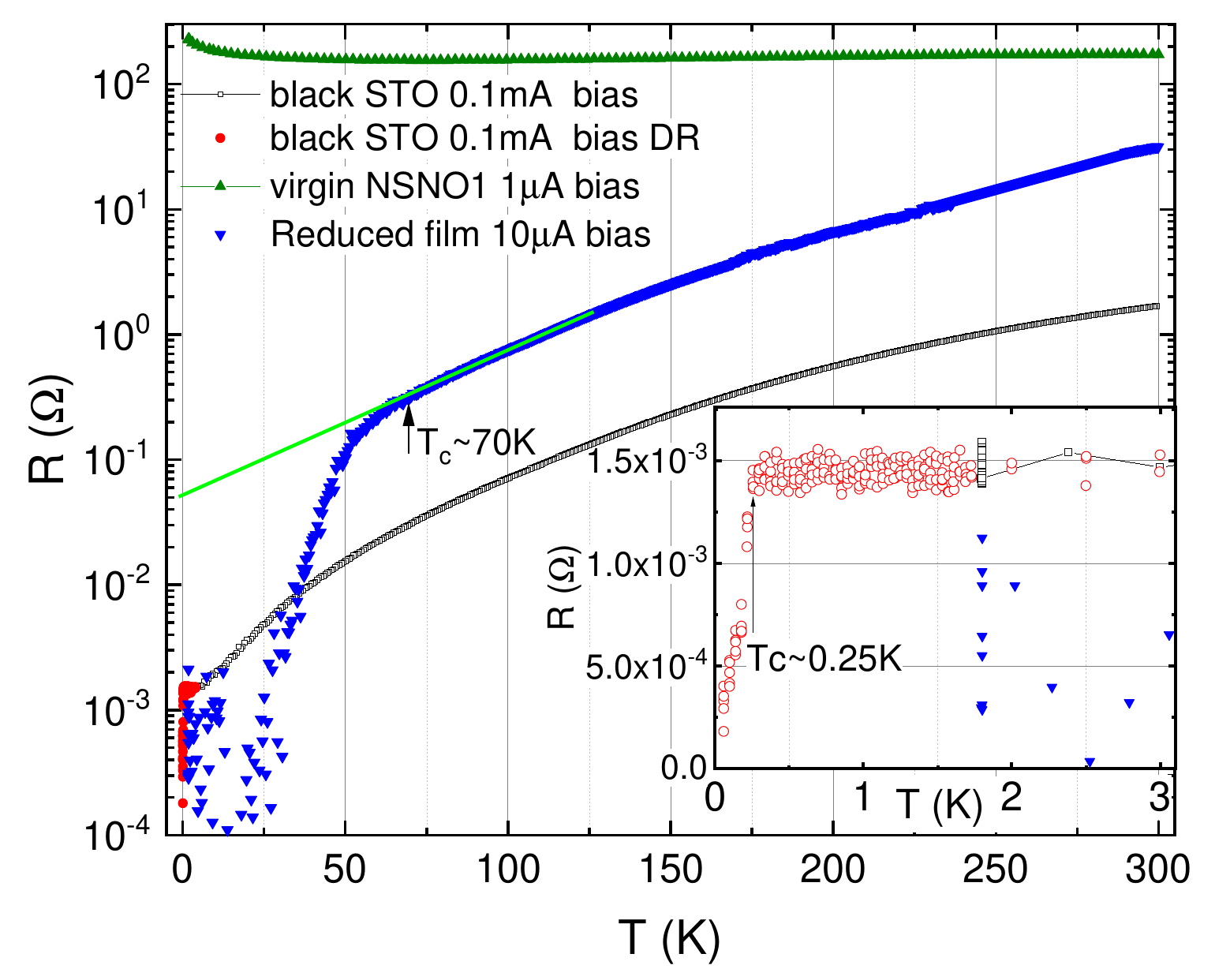}
 \caption{ Resistance versus temperature on a log-linear scale of the virgin NSNO1 and reduced NSNO1red1h films, together with the R vs T of a bare black STO wafer reduced in $\rm CaH_2$ at $320\,^\circ\mathrm{C}$ for 6 h. In the inset,  a zoom-in on the $R$ vs T data of the main panel below 3 K is shown,  where the red circles represent measurements obtained in a dilution refrigerator (DR). The transition to superconductivity at $\sim$0.25 K is due to the highly reduced, optimally doped, black STO wafer. }
\label{fig3}
\end{figure}

Two transitions are seen in Fig. 2 (b). A sharp superconducting one with $\mathrm{T}_\mathrm{c}$ onset at about 7 K, and a more moderate one, clearly showing diamagnetism and possibly superconducting, with $\mathrm{T}_\mathrm{c}$ onset at about 50 K. The sharp low temperature transition is well established by now as originating in superconductivity. This was shown in resistivity and magnetic susceptibility measurements, where in $\rm Nd_{0.8}Sr_{0.2}NiO_2$ films susceptibility drops  of 20\% \cite{Hwang} and 50\%  \cite{LeeHwang} were observed just below the transition, while here we found a comparable magnetization moment drop of 30\%. We believe that the second weaker diamagnetic transition at about 50 K also originates in superconductivity, but in islands here, which leads to its broadening. We stress that in the same kind of films ($\rm Nd_{0.825}Sr_{0.175}NiO_2$ on STO \cite{Saykin-Kapitulnik}), above the low-temperature superconducting transition, magnetization versus temperature measurements under field warming (FW) after ZFC protocol as used here, showed no diamagnetism at all. On the contrary, it showed a monotonous decrease of M with increasing temperature. Furthermore, in bulk samples of the undoped compound NdNiO$_2$ \cite{HaiLin}, magnetic-susceptibility versus temperature measurements also didn't show diamagnetism above the low temperature $\rm T_c$, just a monotonous decrease with increasing T. We are therefore quite confident that the transition we see at about 50 K actually originates in superconductivity. The scenario in which the diamagnetic transition at 50 K originates in puddles of magnetic Nd, Ni or their oxides in our films was ruled out by its absence under the application of a field parallel to the film's surface.  Using scanning SQUID magnetometry, Shi \textit{et al.} found a background of magnetic orders coexisting with superconductivity in infinite-layer nickelate films \cite{Shi}. This phenomenon can also affect our results and will be further discussed in  Section 5 of the supplement.\\

Figure 2 (c) depicts the moments versus temperature at a few low fields, showing that their magnitude is proportional to the fields used, and also that the transition onset temperature is slightly suppressed with increasing field. We point out that the moments in Fig. 2 (b) are negative while in Fig. 2 (c) they are positive. This seemingly puzzling result is due to the magnetic history of the film and substrate preceding the measurements of the moments. The hysteretic moments versus field of Fig. 2 (d) clearly explains this behavior, as a previous history of high negative fields (-5000 Oe here) will yield negative moments at low fields and vice versa. The hysteresis in this figure also suggests the existence of ferromagnetism in the STO substrate, possibly due to magnetic impurities,  with a saturation field of about 800 Oe. This was established by measuring a bare STO wafer (without a film), where results similar to Fig. 2 (d) were found (see Fig. S2 in the supplement). We should note that a more useful magnetic moment unit of $\rm Am^2$ can be used instead of $\rm emu$ given by the MPMS3 in the magnetization data presented here. This would place the signal in the $\rm \sim 10^{-11}Am^2$ range for 1 Oe of applied magnetic field.\\

We now turn to the transport results of this film before and after annealing for 1 h in $\rm CaH_2$. Figure 3 shows the corresponding resistance versus temperature of these films, together with a typical background result of a bare black STO wafer also annealed in $\rm CaH_2$ but for 6 h. The normal resistance at 100 K is $\sim$1 $\Omega$ in the reduced film  and that of the black STO is about an order of magnitude lower. However, we can't calculate the resistivity of our films since both film and substrate are conducting, we don't know how thick the conducting layer in the STO is, and we actually have a bilayer with interactions between the layers. Therefore, comparison with $\rho$ values obtained in previous studies can't be made. A clear resistance drop which starts at about 70 K is seen in the annealed film, with crossing of the black STO curve at $\sim$35 K. Overall this resistance drop is of more than two orders of magnitude down to the noise level below 1 m$\Omega$ at about 25 K. We stress that the noise level will always mask an actual zero resistance, which could be extracted from voltage versus current curves as will be shown later when discussing the results of Figs. 7 and 8. As a stand alone result, the resistance drop in Fig. 3 starting at about 70 K, does not necessarily originate in superconductivity. However, in view of the observations of a diamagnetic transition (as in Fig. 2 (b)) and the paramagnetic Meissner effect (as in Fig. 5 (c)) at the same temperature range of $\sim$50 K, we believe that the resistance drop in Fig. 3  can safely be attributed to a superconductive transition with $\rm T_c$ onset of 50-70 K.\\ 

Our results indicate conduction between superconductive islands below 70 K via inter-islands weak-links, which become superconductive at $\mathrm{T}_\mathrm{c}$ offset of 25 K. We believe this is the highest $\mathrm{T}_\mathrm{c}$ onset reported at ambient pressure in reduced Nickelate films. This transition could originate in any of the low n RP phases (n=2, 3 or 4) as seen in Fig. 1 (f), or in the 112 IL phase, all of which with oxygen deficient stoichiometry.  Figure 3 also shows that besides the fact that the normal black STO resistance is about an order of magnitude lower than that of the Nickelate film, this STO  has a superconducting transition $\mathrm{T}_\mathrm{c}$ at 0.25 K as depicted in the inset to this figure.  This is typical of oxygen deficient, optimally reduced STO as reported in the literature \cite{blackSTO}.

\subsection{Longer reduction time film}

\begin{figure}
 \centering
        \includegraphics[width=1\textwidth]{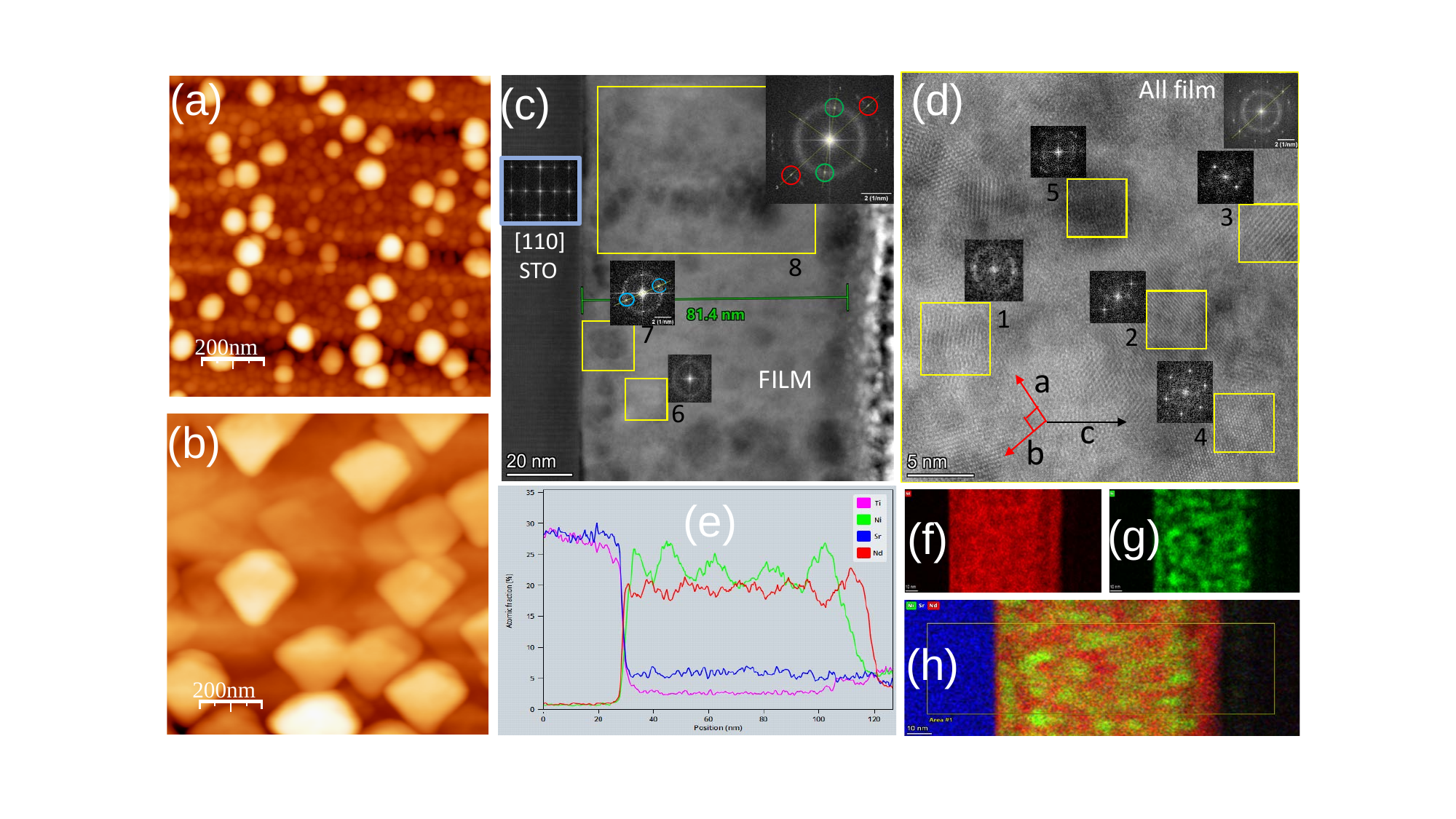}
 \caption{AFM, STEM and EDS results of the 80 nm thick NSNO2 virgin film and the reduced NSNO2red6h film. (a) and (b) show AFM images of the surface morphology of the NSNO2 and NSNO2red6h films, respectively. The square shaped crystallite outgrowths in (b) are likely due to $\rm NdO_x$ in the top Nd-rich layer of this film (see the EDS profile in (e)). (c) and (d) depict STEM images of the 6 h reduced film, where it became polycrystalline. FFT of different regions of the reduced film are shown adjacent to the marked areas involved (the yellow rectangles). More intense diffraction peaks observed in addition to the fainter polycrystalline rings are circled in the FFT of areas 7 and 8 in (c), and will be used in the supplement to identify the phases involved.  (e) depicts EDS data of this film where the net Nd+Sr atomic fraction is equal to that of the net Ni+Ti, except for the top surface layer which is Nd-rich.  (f-h) show atomic distribution maps of the elements in this film, but now the Ni distribution in the film is even less uniform than that seen in Fig. 1 (h). The overall ratio of the [Nd+Sr]/[Ni+Ti] atomic fraction is still around 1, but the fluctuations are about twice as large as in Fig. 1 (g) (about 10 vs 5\%).  }
\label{fig4}
\end{figure}

The strikingly high $\mathrm{T}_\mathrm{c}$ values observed in our reduced film at ambient pressure, compelled us to double check these results and repeat the experiments on a second freshly prepared film. The laser fluence on the target this time was higher  than in the original NSNO1 film (2.5 versus 1.5 J/cm$^2$), resulting in a much thicker film. Figure 4 (a) and (b) depict two AFM images of this 80 nm thick film, where the surface morphology of the virgin film NSNO2 is seen in (a), and that of the reduced film NSNO2red6h in (b) (reduced in $\rm CaH_2$ at $320\,^\circ\mathrm{C}$ for 6 h). One can see many particulates and out-growths of 10-25 nm height on the virgin film, and good crystallization at the background with grain sizes of about 50 nm.   The reduced film shows well developed squarish crystallite outgrowth of about 10 nm height. We identified these cubic-like outgrowths as $\rm NdO_2$ or $\rm Nd_2O_3$ crystallites, as was found from FFT profiles taken on the Nd rich region at the top of this film (see the EDS data in Fig. 4 (e)). \\

HAADF-STEM images of the 4 months old NSNO2red6h film after its transport and magnetic measurements are shown in Fig. 4 (c) and (d), together with FFT images of specific regions of this film. The STEM lamella was prepared by FIB lift-off along the [110] zone axis, as also observed by the corresponding FFT of the STO wafer in (c).  The c-axis in (d) is in the image plane, while the a and b axes are in a plane normal to the image and at $\rm \pm 45^o$ to it. One can see that this film is very disordered and polycrystalline as indicated by the rings observed in most FFT images in (c) and (d), like in powder diffraction. Nevertheless, some areas of the film show stripes at $\rm 0^o$ and $\rm \pm45^o$ to the interface as depicted by regions 1, 2 and 3 in Fig. 4 (d), as well as the hexagon in the [111] direction as seen in area 4. \\

In Figs. S7 and S8 of the supplement we present FFT images and intensity profiles of some of the marked areas in Fig. 4 (c) and (d). The FFT of area 8 in Fig. 4 (c) is reproduced in Fig. S7 (a), and three FFT intensity profiles along the lines marked in it are drawn in Fig. S7 (b). The broad shaded peak in Fig. S7 (b) is the result of the broad polycrystalline ring. The yellow lines in Fig. S7 (a) were drawn via some more intense FFT spots which originate in larger ordered single crystallites in the film, and these gave rise to the narrower peaks in Fig. S7 (b). These peaks were identified as due to the 214 and 4310 RP phases. In the darker area 7 in Fig. 4 (c), prominent FFT spots were observed (circled in blue). A line profile through them was analyzed in Fig. S8 (a) to reveal that the corresponding crystallites belong to the 327 RP phase.  Similarly, FFT images and profiles through the diagonals shown are plotted in Fig. S8 (b-d), for the whole film in Fig. 4 (d) and in areas 2 and 3 in it. Analysis of all the profiles points to the presence of the n=2, 3 and 4 RP phases in this NSNO2red6h film. Yet, the FFT ring could contain other phases like the 113  RP or the 112 IL phases that we couldn't resolve. \\

\begin{figure}
 \centering
        \includegraphics[width=1\textwidth]{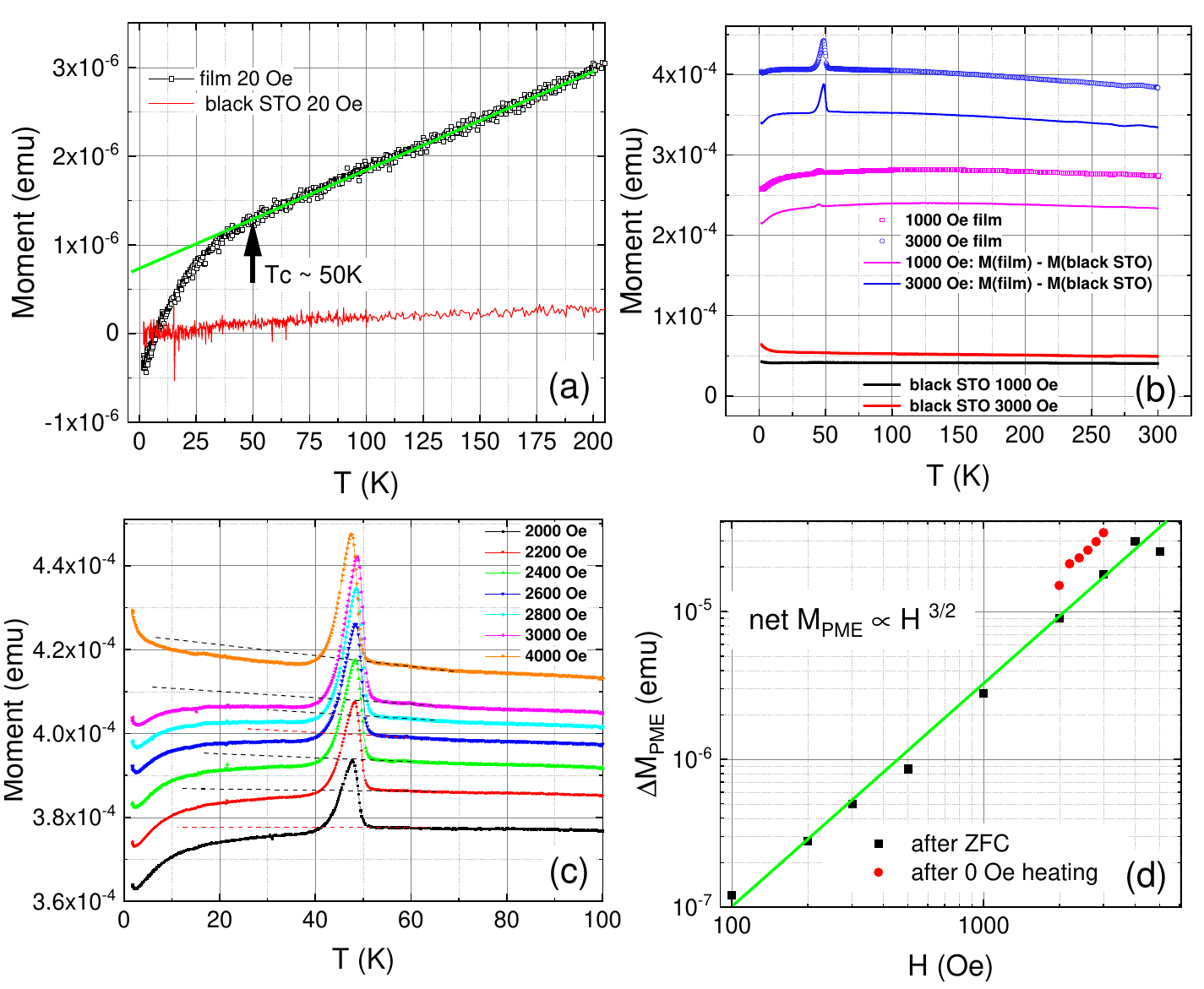}
 \caption{Magnetic moment results. (a) shows the Meissner moment versus temperature of the NSNO2red6h film and of a bare black STO wafer under the same reduction conditions. The latter represents the small background contribution of the substrate to the moment of the film. The moments of this film and its bare substrate are depicted in (b) under 1000 and 3000 Oe, together with the net moments of the film after subtraction of the substrate contribution. (c) shows consecutive moment measurement results versus temperature and under different fields, where the paramagnetic Meissner  effect (PME) peak is robust. In (d) we plot the net PME peak magnitude versus field, where the line represent a power-low behavior with exponent of 3/2. }
\label{fig5}
\end{figure}

Magnetization measurements versus temperature of the oxygen deficient NSNO2red6h sample are shown in Fig. 5 (a), (b) and (c). In this set of measurements, the standard Meissner protocol of zero field cooling, followed by applying the field at 1.8 K and measuring the moments while heating, was followed by an  oscillo-ative field zeroing at 300 K, to ensure that the field was actually close to zero in the next ZFC (and for the initial cooldown, even less than 0.02 Oe). As can be seen in (a), a Meissner effect is apparent, above a small magnetic background from the STO substrate. The transition again isn't sharp, yet it is clear that the transition temperature is at about 50 K. In Fig. 5 (b) moment results are shown for two higher fields with their corresponding black STO background. Also shown are the net film signals after subtraction of their STO backgrounds. We note that this time, both film and bare STO wafer went through the same 6 h annealing in $\rm CaH_2$. The new feature observed under these fields is an emerging positive peak just below 50 K. This peak can be attributed to the paramagnetic Meissner effect (PME), which generally appears just below the superconducting transition temperature \cite{Wohlleben}.\\

The appearance of the PME can be explained theoretically as due to the presence of an odd-frequency order parameter, the existence of SN or SF junctions in our films, or a giant vortex state in them \cite{Linder,Koblischka,Moshchalkov}. Junctions are quite obviously present in our films, as depicted in the schematic diagram of the inset to Fig. 6. This will be elaborated on in more details in the discussion section. More systematic measurements were carried out under fields of 2000 to 4000 Oe as shown in Fig. 5 (c), where the giant PME peaks are clearly larger than the magnitude of the whole Meissner effect drop with decreasing temperature.  We note that the magnitude of these Meissner drops decreases with increasing field, and disappears at 4000 Oe, though the moment at temperatures above the PME peaks are always higher than just below it (see the dashed lines). The main point though of the observation of a giant PME at  $\sim$48 K in our films, is to further support the existence of superconductivity with $\mathrm{T}_\mathrm{c}$ slightly above it at 50 K. In Fig. 5 (d) we plot on a log-log scale the net PME peak magnitude versus field and find a power law behavior with an exponent of 3/2. We know of no theory to explain this behavior at the present time. \\

Transport results of $R$ versus T of NSNO2red6h are shown in Fig. 6 on a linear scale this time, together with the black STO background (times 10). $\mathrm{T}_\mathrm{c}$ onset here is again at 70 K as in NSNO1red1h of Fig. 3. In addition to the results obtained by the PPMS in Fig. 6, data extracted from the voltage versus current curves measured by a Keithley 2450 is also given here. The latter represent more accurate data at low temperature as discussed in detail in Figs. 7 and 8. By taking only the flat superconductive part of the V vs I curves of Fig. 7, one gets supercurrents. The low bias R values calculated from the linear part of these VICs are plotted by the red circles in Fig. 6. This shows a near-coincidence with the PPMS results above 25 K, with a drop to zero resistance below about 20 K. Comparing the results of Fig. 6 and Fig. 3, one sees that the R values drop to the noise level at $\sim$20 K and $\sim$25 K, respectively.  We thus conclude that the basic Meissner and $\mathrm{T}_\mathrm{c}$ properties of both NSNO1red1h and NSNO2red6h films are quite similar. The differences in the observation of the giant PME peak and asymmetric diode effect in NSNO2red6h might be attributed to the longer annealing duration of this film. In the supplement in sections 3 and 4, we present data of two more films we measured, of 10 and 20 nm thickness, with varying annealing times, which support the $\rm T_c>$50 K enhancement  result here.  \\

\begin{figure}
 \centering
        \includegraphics[width=1\textwidth]{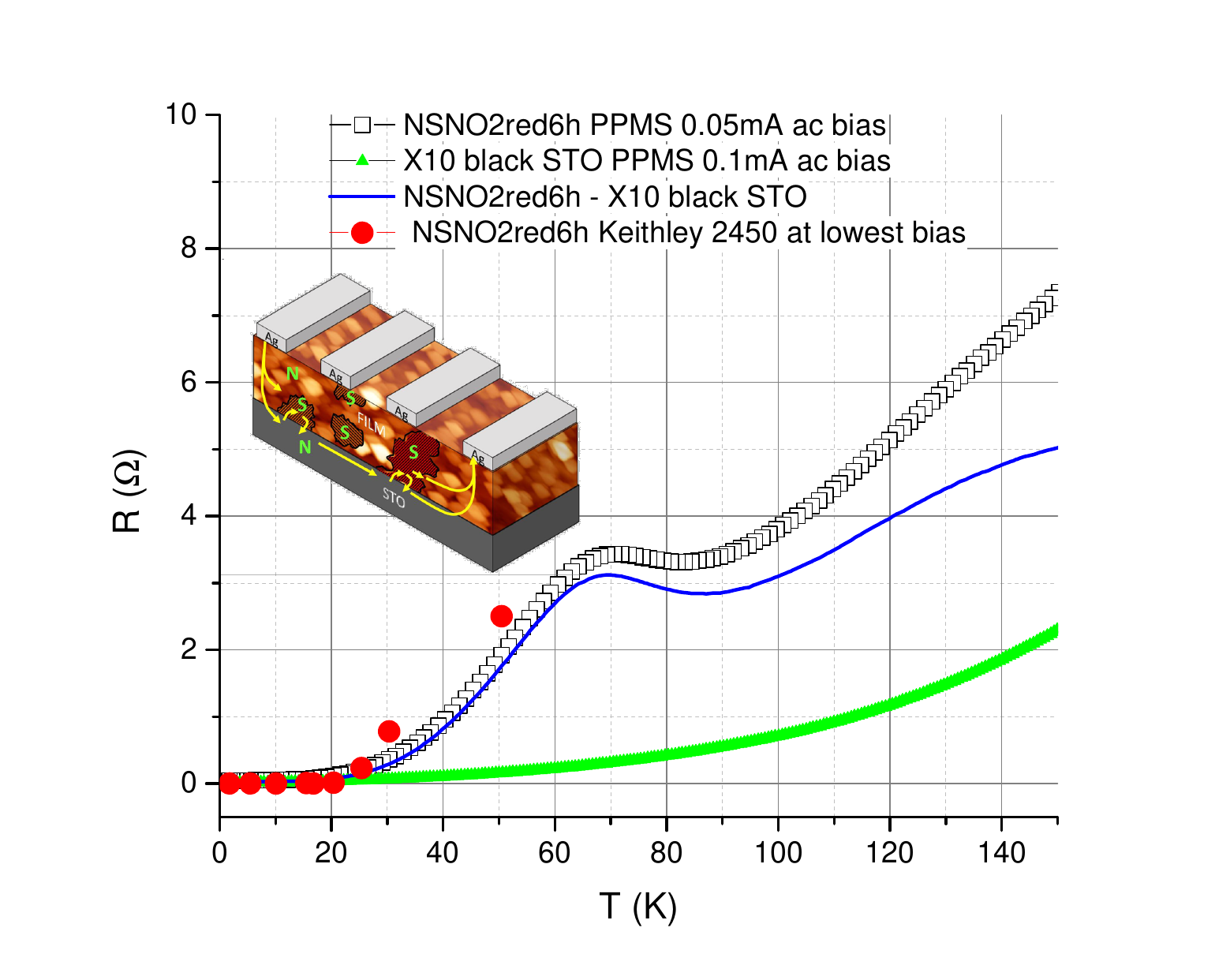}
 \caption{Resistance versus temperature on a linear scale of the reduced NSNO2red6h film (black squares), together with the R vs T of a bare black STO wafer times 10 (green triangled), both measured using the PPMS system and its software. Also shown is the net film resistance obtained by subtracting the black STO background (blue curve). The red circles represent resistance values measured  from the low bias IVCs by a Keithley 2450 source-meter, by taking into account only the minimal resistance parts of the IVCs and avoiding the resistance parts due to the diode effect (below about 20 K). The inset shows  a schematic drawing of the sample, bias current flow directions and the contacts geometry, where the shaded areas represent the superconducting islands (S). }
\label{fig6}
\end{figure}

Next we present the diode effect and supercurrent results of NSNO2red6h as shown in Figs. 7 and 8. We start with Fig. 7 (a) and (b) where the ramping up of the current is separated from the ramping down by their respective colors (black and green). One can see that the V vs I curves here have two typical kinds of noise. The standard voltage noise with peak to peak amplitude of about 5 $\mu$V, and a telegraph noise jumps of about 7 $\mu$V. These noises are almost averaged out to about 12 $\mu$V when sampling on a larger number of points (2000 vs 1000 points in (b) and (a), respectively). A remarkable feature in Fig. 7 is the observation of highly asymmetric curves which indicates a clear diode effect. We extract the supercurrents from the flat horizontal parts of these curves as depicted by the blue lines in Fig. 7. For instance, at 12 K in (c) one can see a small flat part with still asymmetric resistances on both its sides. This resistance asymmetry is almost gone at 20 K as seen in (d). We note that the whole V vs I curves here are shifted to small positive voltages. The shift is 0.35 mV at 1.8 K, while with increasing temperature it decreases to 0.22 mV at 20 K. This voltage shift behavior vs temperature can be better seen in the current vs voltage curves in Fig. 8 (c), where at 50 K it completely disappears.  These voltage shifts are unclear to us at the present time, but they could arise from thermo-electric effects between the voltage contacts. \\

\begin{figure}
 \centering
        \includegraphics[width=1\textwidth]{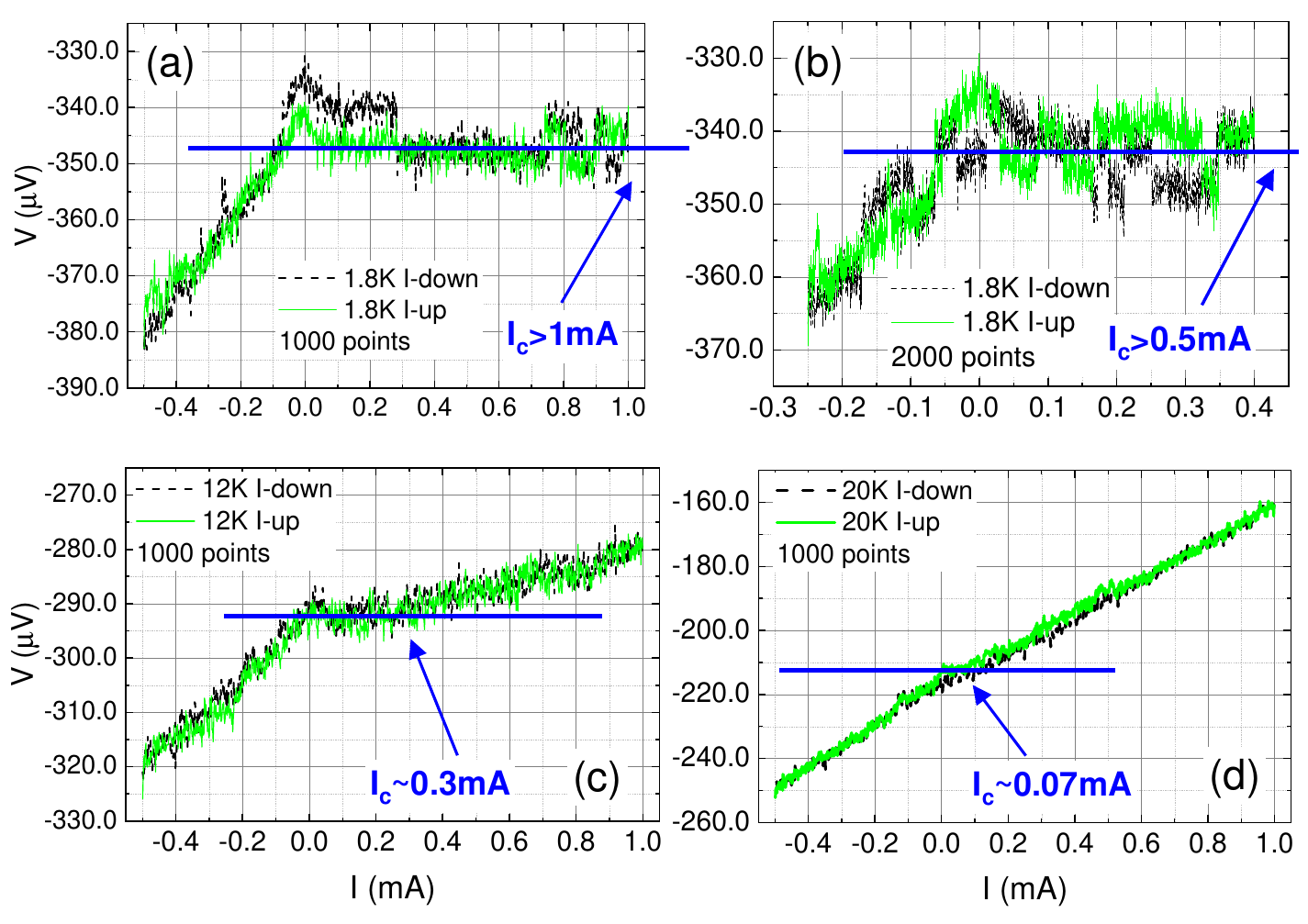}
 \caption{Voltage versus current curves of NSNO2red6h with one week aged Ag contacts as appeared on the monitor of the Keithley 2450. No averaging was used in these measurements. (a) and (b) were taken at 1.8 K, (c) at 12 K and (d) at 20 K. Up-current scans are plotted in green, while down-current scans are plotted in black. Running voltage noise level at 1.8 K is about 5 $\mu$V and telegraph noise jumps are of about 7 $\mu$V.  }
\label{fig7}
\end{figure}

Figure 8 presents the more standard current versus voltage curves (IVCs) and the supercurrents versus temperature. The main differences between Fig. 7 and Fig. 8 are that the former represents raw data measured with one-week-old Ag-paste contacts, while the latter shows smoothed data measured on freshly prepared Ag-paste contacts. Fig. 8 (a) and (b) show how the $\pm$2.5 $\mu$V criterion to determine the critical current was used (see the shaded rectangles). It also shows that while in (b) the diode effect is fully asymmetric (resistance starts immediately above zero bias), in (a) the asymmetry shows both positive and negative supercurrents ($\rm I_c^+$ and $\rm I_c^-$, respectively). In Fig. 8 (c), a few IVCs are plotted, and the resulting combined supercurrents $\rm I_c^++|I_c^-|$ as extracted from these and more IVCs are given versus temperature in (d). We note that below about 5 K, $\rm I_c^++|I_c^-|$ in Fig. 8 (d) should actually be higher than 1 mA as can be deduced from Fig. 7 (a) which was obtained with different contacts and without averaging.

\begin{figure}
 \centering
        \includegraphics[width=1\textwidth]{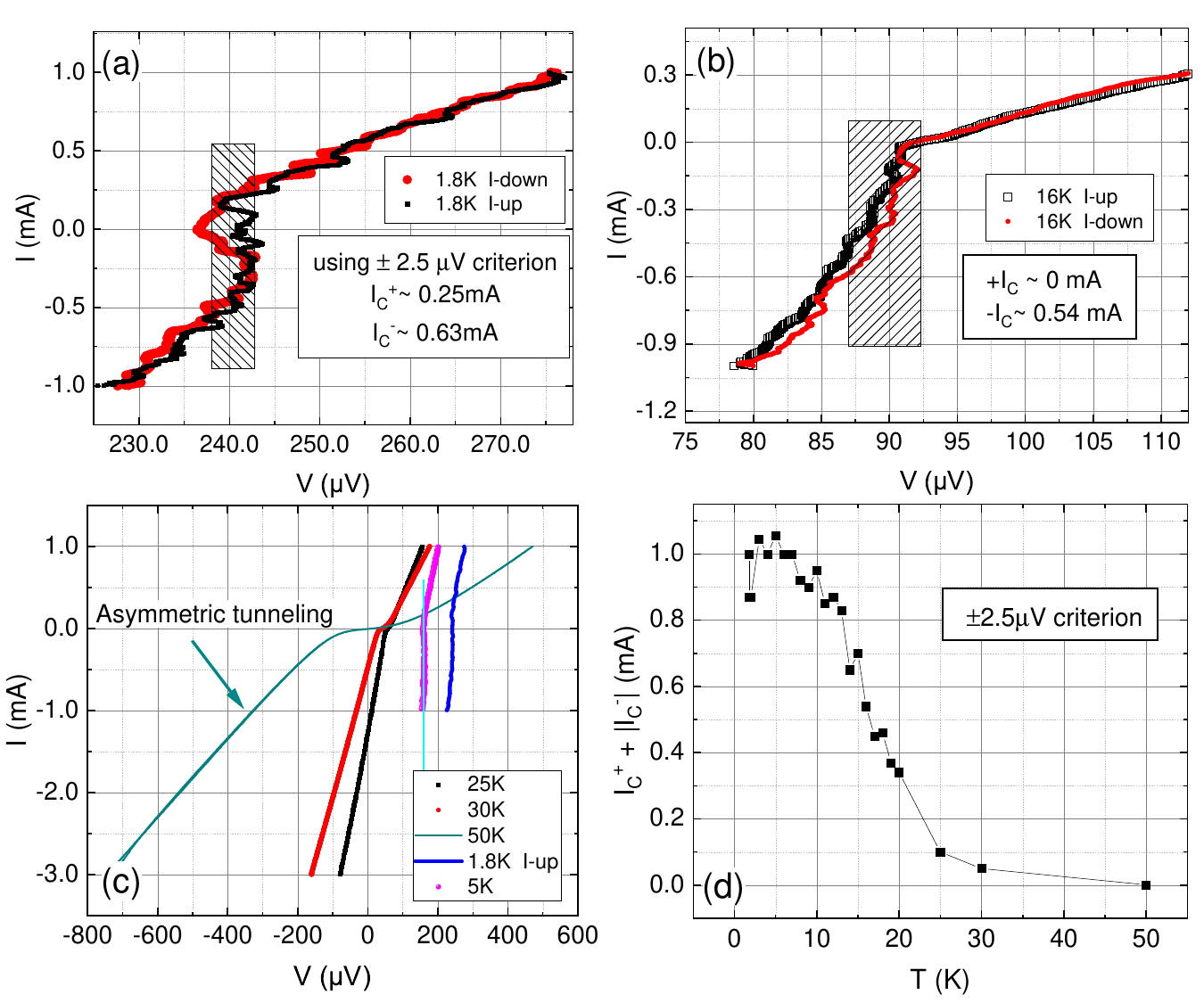}
 \caption{Current versus voltage curves (IVCs) of the NSNO2red6h film are depicted in (a)-(c) at different temperatures, showing the asymmetric diode effect. Note the reversed polarity at low bias of these curves (more resistive at positive bias) as compared to that of Fig. 7 (more resistive at negative bias). The width of the shaded boxes in (a) and (b) represent the   $\pm$2.5 $\mu$V criterion we used to determine the critical currents which are plotted in (d) versus temperature. Note that the S-shaped band near zero bias of the red curve in (a) is due to averaging over a telegraph-noise jump which isn't included in our voltage criterion (see text).  (c) shows that at 50 K there is no $I_c$ and the IVC is still asymmetric and clearly tunneling-like. It also shows the decrease with temperature of the small voltage shift from zero at zero bias.  }
\label{fig8}
\end{figure}

\section{Discussion}

In the present study, experimental evidence for three novel phenomena in nickelate films was found. $\mathrm{T}_\mathrm{c}$ enhancement to 50-70 K at ambient pressure, a giant paramagnetic Meissner effect peak around 48 K, and a polarity-reversible diode effect. To the best of our knowledge, $\mathrm{T}_\mathrm{c}$ above 50 K as found here is the highest reported in reduced nickelate films at ambient pressure. Since this effect seems to be strongly dependent on the highly reduced STO substrate, we decided to repeat the experiment on a different new film but with a longer annealing time. The high $\mathrm{T}_\mathrm{c}$ observed was well reproduced in this new film, but in addition giant PME peak and diode effects were discovered. In the following, we discuss these three phenomena.

\subsection{$\rm T_c$ enhancement}
Ever since the discovery of superconductivity in reduced nickelate thin films \cite{Hwang}, the quest to increase their $\mathrm{T}_\mathrm{c}$ has been ongoing. By better controlling the deposition, composition and annealing in $\rm CaH_2$ conditions, success was achieved but strains with the substrate on which the films were deposited, played a major role in obtaining the highest $\mathrm{T}_\mathrm{c}$ value \cite{Segedin2023}. Instead of the commonly used STO substrate, the use of $\rm NdGaO_3$ substrates yielded the highest $\mathrm{T}_\mathrm{c}$ values so far (25 and 38 K) \cite{FreeStanding,ChowAriando}. In our study, we still use the STO substrate, but in a highly reduced and conducting state and this seems to be the essential ingredient in pushing $\mathrm{T}_\mathrm{c}$ up. The lattice constant of reduced and normal STO are almost the same \cite{SameLatticeConstant}, therefore the increased $\mathrm{T}_\mathrm{c}$ doesn't seem to originate in additional strains in our case. This conclusion is also strengthened by the fact that the present films in which $\mathrm{T}_\mathrm{c}$ was enhanced are 30-80 nm thick, too thick to be affected by strain from the substrate, unless this is an interface effect. The later might actually be the case, as can be deduced from the results in Figs. S3 and S4 of the supplement, obtained in 20 and 10 nm thick films.\\

If in fact, the film's layers in close proximity to the substrate  are responsible for the elevated $T_c $ observed, then the top layers of the film serve to protect the bottom layers from ambient effects, and the reduced substrate serves as an effective oxygen getter, keeping the film well reduced over time. We noticed aging effects of our films in which over a few months became very resistive, and the STO substrates lost their blackness  and returned to their normal whitish transparent color. This obviously originated in the reduced STO absorbing oxygen from the air, indicating its strong efficiency as a getter even at room temperature. \\

As discussed before the STEM and EDS results of Figs. 1 and 4 indicate that the phases involved in the present study are the various RP phases with n=2, 3, 4 and n$\rightarrow\infty$ and the IL phase. Though equal average atomic fractions of [Nd+Sr] and [Ni+Ti] were found in both NSNO2red1.5h and NSNO2red6h films, the Ni aggregation in them renders them inhomogeneous and allows for the various RP phases to exist in different regions of these films. Possibly, the well oxygenated 327 phase that showed the highest $\rm T_c$ under pressure ($\sim$80 K) and $\sim$40 K at ambient pressure is also responsible to the elevated $\rm T_c$ that we observed in the present study. Since our films are highly reduced, it could be that similar to the cuprates, they are like the electron doped superconductors. The XRD in Fig. 2 (a) points to the presence of the infinite-layer nickelate phase, that was reported already in the pioneering study in which superconductivity in the nickelates was first observed \cite{Hwang}. Figure 2 (b) clearly shows that we have two Meissner transition temperatures in NSNO1red1h which was reduced for 1 h. In Fig. 5 (a), where the low $\mathrm{T}_\mathrm{c}$ transition at 7 K is absent, this is due to the 6 h long annealing of NSNO2red6h, since in another quarter of the NSNO2 film reduced for 1.5 h to produce NSNO2red1.5h, the 7 K transition was clearly seen (not shown). The XRD of NSNO2red6h film showed no film peaks besides  a small remanence "knee" of the (002) peak of the perovskite phase. This is explained in section 1 and Fig. S1 of the supplement. 

\subsection{ Meissner and Paramagnetic-Meissner effects}

Observation of the Meissner effect is a clear signature of superconductivity. It can be seen in Fig. 2 (b) and Fig. 5 (a) where zero field cooling was followed by heating under 20 Oe. The former shows the low temperature superconducting transition of the $\rm Nd_{0.8}Sr_{0.2}NiO_2$ infinite-layer phase with $T_c\sim$7 K, together with the $T_c\sim$50 K transition. This was obtained in a film that was reduced in $\rm CaH_2$  for 1 h only. When the reduction time was increased to 6 h on a different film as in Fig. 5, the low-temperature transition disappeared and only the high-temperature one persisted. This result could originate in the higher disorder of the longer annealed film as seen in the STEM images of Fig. 4 (c) and (d). Further support for the notion that the $\rm T_c\sim$50 K transition originates in superconductivity, is given in Fig. 5 (c) where a giant paramagnetic Meissner effect peak appears at $\sim$48 K. Such a peak signifies the presence of superconductivity with a transition temperature slightly above it. \\ 

The PME peak can be attributed to odd-frequency order parameter which breaks inversion and time reversal symmetries \cite{Linder}. It could also originate in SN or SF junctions in our film \cite{Koblischka}, or develop from the formation of a giant vortex state in them \cite{Moshchalkov}. We can not comment on the order parameter symmetry in our films, since we do not have scanning tunneling conductance spectroscopy results on them. It would be interesting to have such data, to prove or refute the odd-frequency nature of the order parameter. The presence of NS or NF junctions in our films is easier to envision from the inset to Fig. 6, where such junctions form at the interface of the superconducting islands and the STO wafer, or between the islands themselves. The vortex state scenario in our films is quite possible, in particular since the robust PME peaks are observed at quite high fields (2000-4000 Oe). Non-equilibrium vortex trapping due to pinning sites or surface barriers could result in delayed vortex relaxations giving rise to a paramagnetic response. Such a PME peak, but of a much smaller magnitude, was observed in cuprates under fields of a few Oe \cite{Reidling}.  
 
\subsection{Diode effect}

Another interesting phenomenon observed in our films is the diode effect as presented in Figs. 7 and 8. The best way to observe this effect is via the voltage versus current curves (VICs) as depicted in Fig. 7 for a few temperatures. In general, a superconducting diode effect (SDE) can be observed in hysteretic (or nonreciprocal) superconductive Josephson junctions, where supercurrents with increasing and decreasing bias currents are different $I_c^+ \ne |I_c^-|$. Then, when using a modulated bias current with amplitude in between these two supercurrents, the system will alternate between resistive and non-resistive states, which is the basic rectifying diode effect \cite{Qi}. Here, in Fig. 7 (a) to (c) one can see \textit{fully polarized} VICs, where on negative bias the films are resistive, while on positive bias they are superconducting. In (d), very close to $\mathrm{T}_\mathrm{c}$ of zero resistance,  the VIC is almost Ohmic all the way with only a tiny supercurrent part. We emphasis that there is almost no hysteresis in our case, and our SDE is more like the diode effect seen in semi-conductors but with a zero resistance in the more conductive branch. Theoretically, the SDE originates in materials with broken time-reversal and inversion symmetries \cite{HeTanaka,Bergeret,Yuan-Fu,Ando}. Most of these studies assumed the presence of an external magnetic field, but some discussed also field-free SDE \cite{Qi}. Internal Zemman fields and disorder in nonreciprocal materials with strong spin-orbit or Rashba interactions, can also facilitate field-free SDE \cite{Bergeret}.\\

To our surprise, the observed SDE was easily polarity reversed as can be seen in Fig. 8 (a) to (c). This was found on the \textit{same} film as in Fig. 7, but with newly prepared contacts with obviously a slightly different geometry. Mori et al. showed theoretically that by changing the distance between the superconducting electrodes, the SDE polarity could be reversed and that this effect is periodical in distance \cite{Mori}. So in our case, the superconducting islands or domains can be considered as S-electrodes, and the black STO as the connecting normal metal (N-electrode) in the Mori et al. scenario. It should be noted that as long as there is no perculative supercurrent in the nickelate film (between $\mathrm{T}_\mathrm{c}$ onset and $\mathrm{T}_\mathrm{c}$(R=0)), the bias current will mostly flow via the highly conductive normal STO wafer, with occasional shunts through adjacent superconducting islands, as depicted in the schematic drawing of the inset to Fig. 6.  

\subsection{Supercurrents}

Figures 7 and 8 were used to determine the supercurrents versus temperature in our film, as depicted in Fig. 8 (d). Due to the asymmetric VICs with different $I_c^+$ and $|I_c^-|$, we determined the combined supercurrents as $I_c^++|I_c^-|$. In the symmetric VIC case, it would amount to 2$I_c$, while in the fully Asymmetric case (as in Fig. 7), it would be the standard $I_c$. We used a $\pm2.5\,\mu$V criterion to asses the combined supercurrent. This criterion can be inferred from the noise of the VICs of Fig. 7 (a) and (b) at 1.8 K. The uninterrupted trace voltage noise is about 5 $\mu$V (2$\sigma_1$), while the occasional telegraph-jumps are about 7 $\mu$V. Thus, the overall noise is about 12 $\mu$V (2$\sigma_2)$. Disregarding the telegraph-noise which is less prevalent at higher temperatures, we end up with the $\sigma_1=2.5\,\mu$V criterion. Fig. 8 (d) shows that the supercurrent in our film decreases significantly at about 25-30 K, and is zero at 50 K. Therefore, much optimization work is needed to push the supercurrent to a higher temperature, closer to the $\mathrm{T}_\mathrm{c}$ onset at 50-70 K. Needless to say that identifying the mechanism responsible for this high $\mathrm{T}_\mathrm{c}$ onset would help attain this goal. Finally, we stress again that the present high $\mathrm{T}_\mathrm{c}$ results are different from those obtained in the well oxygenated $\rm La_3Ni_2O_7$ phase \cite{HighPressure,Ko,Zhou2024,Liu,Flavenot}, in that our nickelate films are highly oxygen deficient and strongly affected by the highly reduced black STO substrate.  

\section{Conclusions}

In this study, we demonstrated three noteworthy phenomena in oxygen-deficient nickelate thin films on highly reduced black STO substrates: A $\mathrm{T}_\mathrm{c}$ onset enhancement to 50-70 K, a giant paramagnetic Meissner effect peak, and a highly asymmetric nonreciprocal superconductive diode effect. All of these results were obtained on highly reduced and highly conducting black STO substrates, which are essential for the present observations. The phase responsible for these effects in our multi phase films could not be singled out, and one possibility is that it arises from an electron doped 327 phase like in the electron doped cuprates. Further research is needed to identify the exact mechanism at the origin of these remarkable phenomena in nickelate films. Finally, from the point of view of practical applications, the presently observed fully polarized SDE could be developed into a desirable device, while the highly reduced black STO could be used as a substrate for a broad variety of films that need permanent reduction.

\ack{We are grateful to Amit Kanigel for useful discussions. We thank Yaron Kauffmann, Galit Atiya and Michael Kalina from the MIKA facility at the Technion, and George Levi from Tel Aviv University, for performing the STEM and EDS measurements.}

\roles{Anna Eyal carried out the measurements, Gad Koren prepared the films, and both analyzed the data and written the paper. }

\appendix
\renewcommand{\appendixname}{}

%command (\verb+\appendix*+ in the place of \verb+\appendix+).

\suppdata{}

\subsection{XRD of the 80 and 30 nm thick films}

Figure S1 (a) depicts XRD spectra of the NSNO2red6h film. Clearly, if there are any film peaks, they are within the noise level of this measurement, or affected by the strong (n00) peaks of the STO substrate. (b) shows a smaller XRD range near the (200) peak of the STO substrate, where results of the virgin films NSNO1 and NSNO2 together with the annealed film NSNO2red6h are plotted. Both virgin films before the annealing have prominent (002) peaks of the perovskite Nd$_{0.8}$Sr$_{0.2}$NiO$_3$ phase. After 6 h annealing in $\rm CaH_2$ at $320\,^\circ\mathrm{C}$ of the NSNO2 film that produced the NSNO2red6h film, the (002) film peak almost disappeared and there is still no clear trace of any other film peaks besides those of the STO wafer. The "knee" at 48$^0$ is apparently due to remanence of the (002) film peak of the 113 phase. 

\setcounter{figure}{0} % Reset the figure counter
\renewcommand{\thefigure}{S\arabic{figure}}

\begin{figure}
 \centering
        \includegraphics[width=1\textwidth]{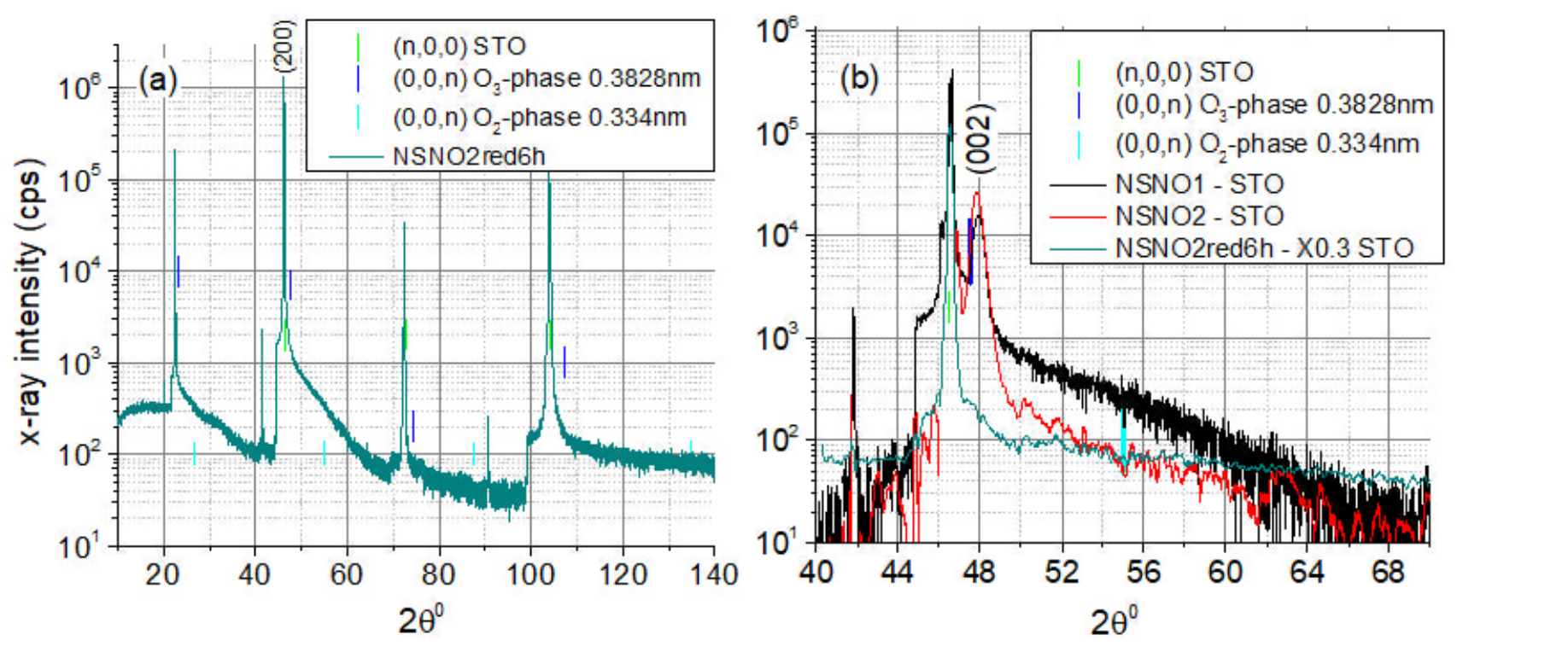}
 \caption{ (a) Full range XRD of the NSNO2red6h film. No apparent film peaks are seen here due to the polycrystalline nature of this film as shown in Fig. 4 (c) and (d), and the high noise level of the XRD system.  (b) XRD of the NSNO1, NSNO2 and NSNO2red6h films near the (200) peak of the (100) STO wafer.  }
\label{figS1}
\end{figure}

\subsection{M vs H of a bare STO wafer}

Figure S2 shows the magnetic moment at 1.8 K and at 100 K versus perpendicular magnetic field of a bare (100) STO wafer without a nickelate film and without annealing in $\rm CaH_2$. The field was cycled from 0 Oe to $\pm$5000 Oe and back to 0 Oe. The data sets were taken after ZFC. The saturation field of $\sim$750 Oe at 1.8 K was reduced to $\sim$550 Oe at 100 K. Clearly, the bare STO substrate behaves like a weak ferromagnet, possibly due to magnetic impurities in it. Because the ferromagnetic component of the magnetization versus H here is very small compared to the diamagnetic component, it has a negligible effect on it. The response at 1.8 K in this figure is similar to that of the reduced NSNO1red1h sample of Fig. 2 (d) of the main article.

\renewcommand{\thefigure}{S\arabic{figure}}

\begin{figure}
 \centering
        \includegraphics[width=1\textwidth]{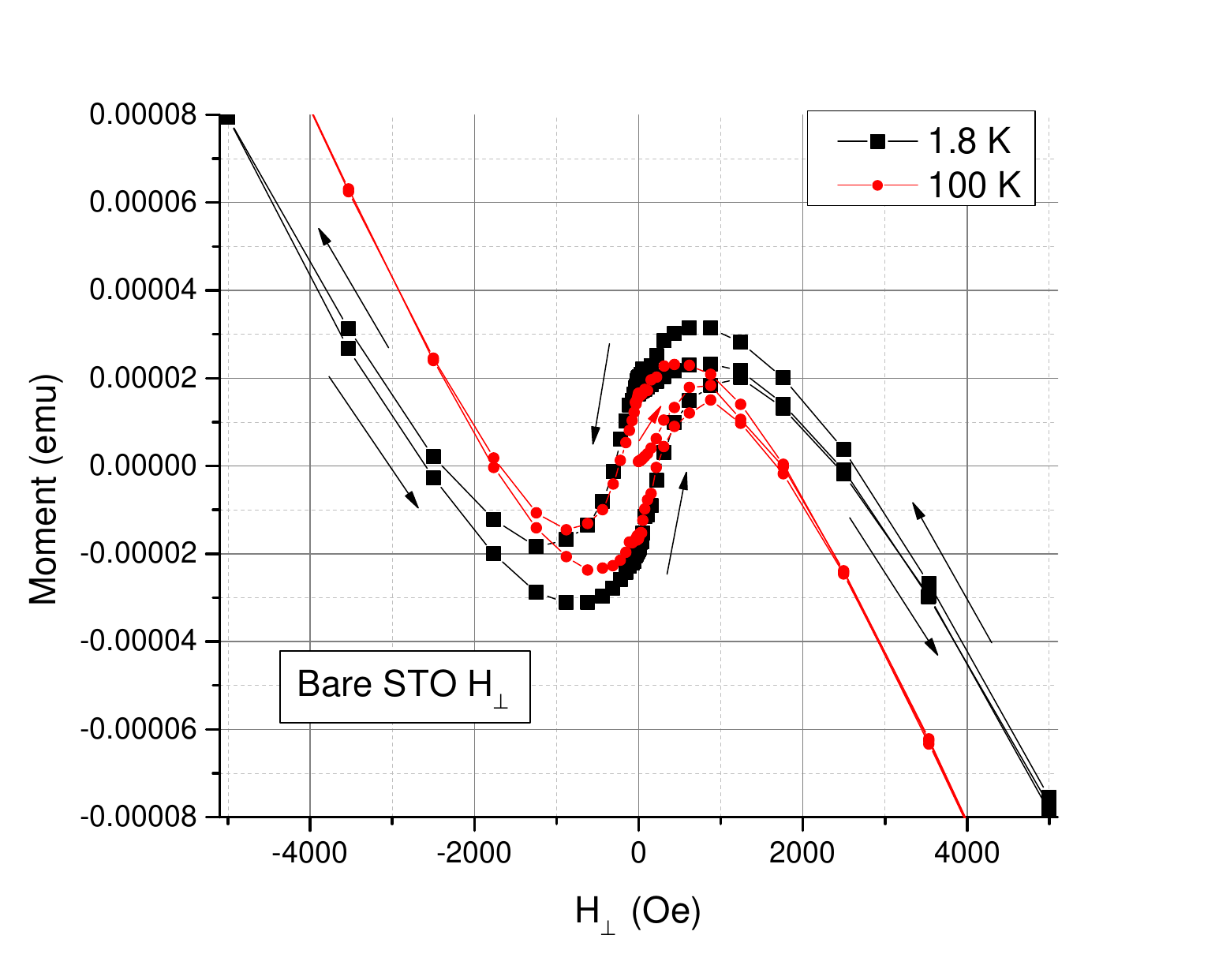}
 \caption{Moment versus perpendicular magnetic field of a bare STO wafer at 1.8 and 100 K. The arrows show the magnetic field cycling sequence starting and ending at 0 Oe.   }
\label{figS2}
\end{figure}

\subsection{$T_c\sim$50 K of a 20 nm thick film}

\renewcommand{\thefigure}{S\arabic{figure}}

\begin{figure}
 \centering
        \includegraphics[width=1\textwidth]{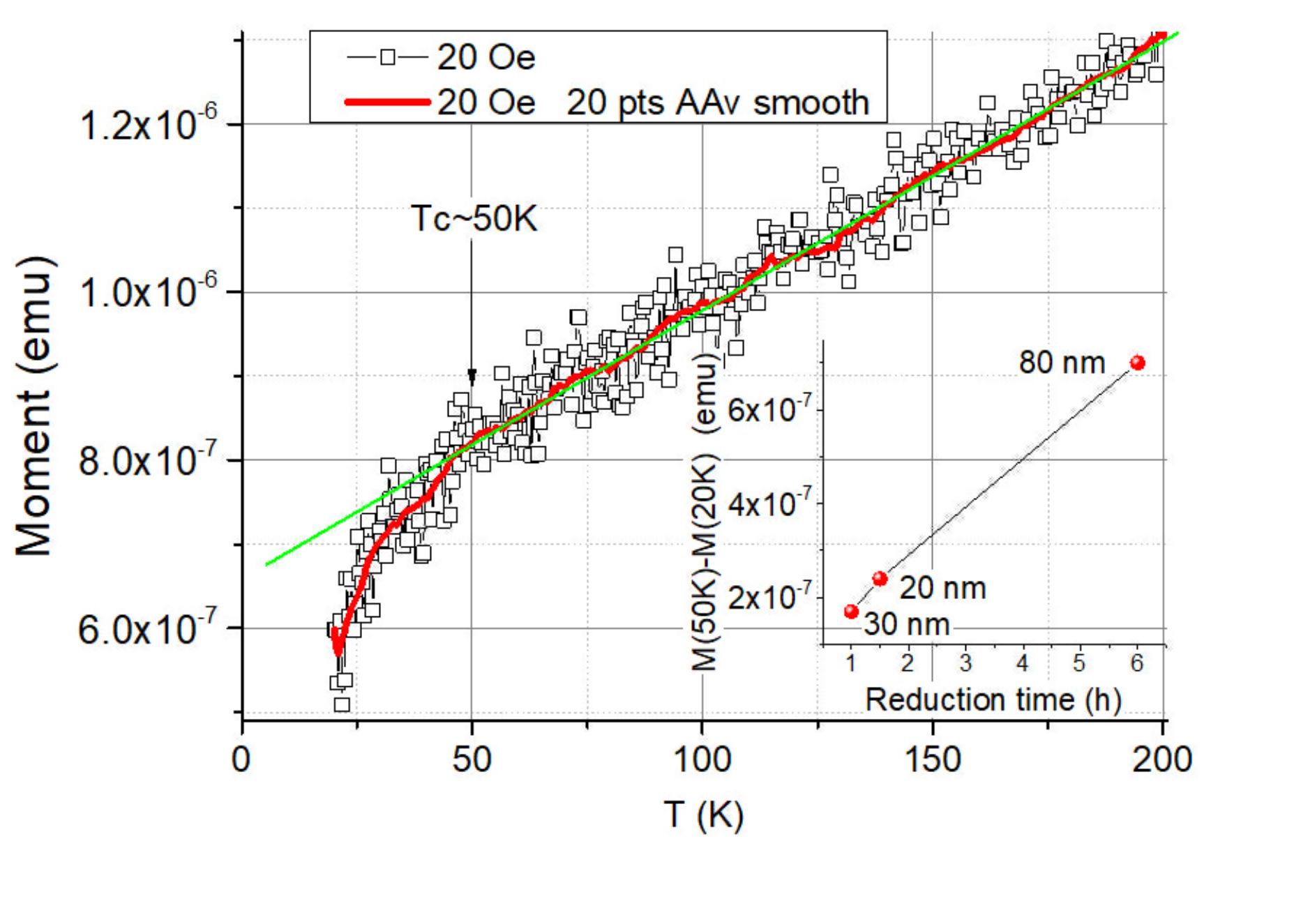}
 \caption{Moment versus temperature of a 20 nm thick NSNO3red1.5h film measured under 20 Oe field. This film was reduced for 1.5 h in $\rm CaH_2$ at $320\,^\circ\mathrm{C}$ under pumping at $\sim10^{-5}$ Torr. The measurement was carried out under heating from 20 K after field cycling to $\pm$5000 Oe at 1.8 K. The inset shows the net moments between 20 and 50 K of three films (NSNO1red1h, NSNO3red1.5h and NSNO2red6h of thicknesses 30, 20 and 80 nm, respectively) versus their reduction time.   }
\label{figS3}
\end{figure}

\renewcommand{\thefigure}{S\arabic{figure}}

\begin{figure}
 \centering
        \includegraphics[width=1\textwidth]{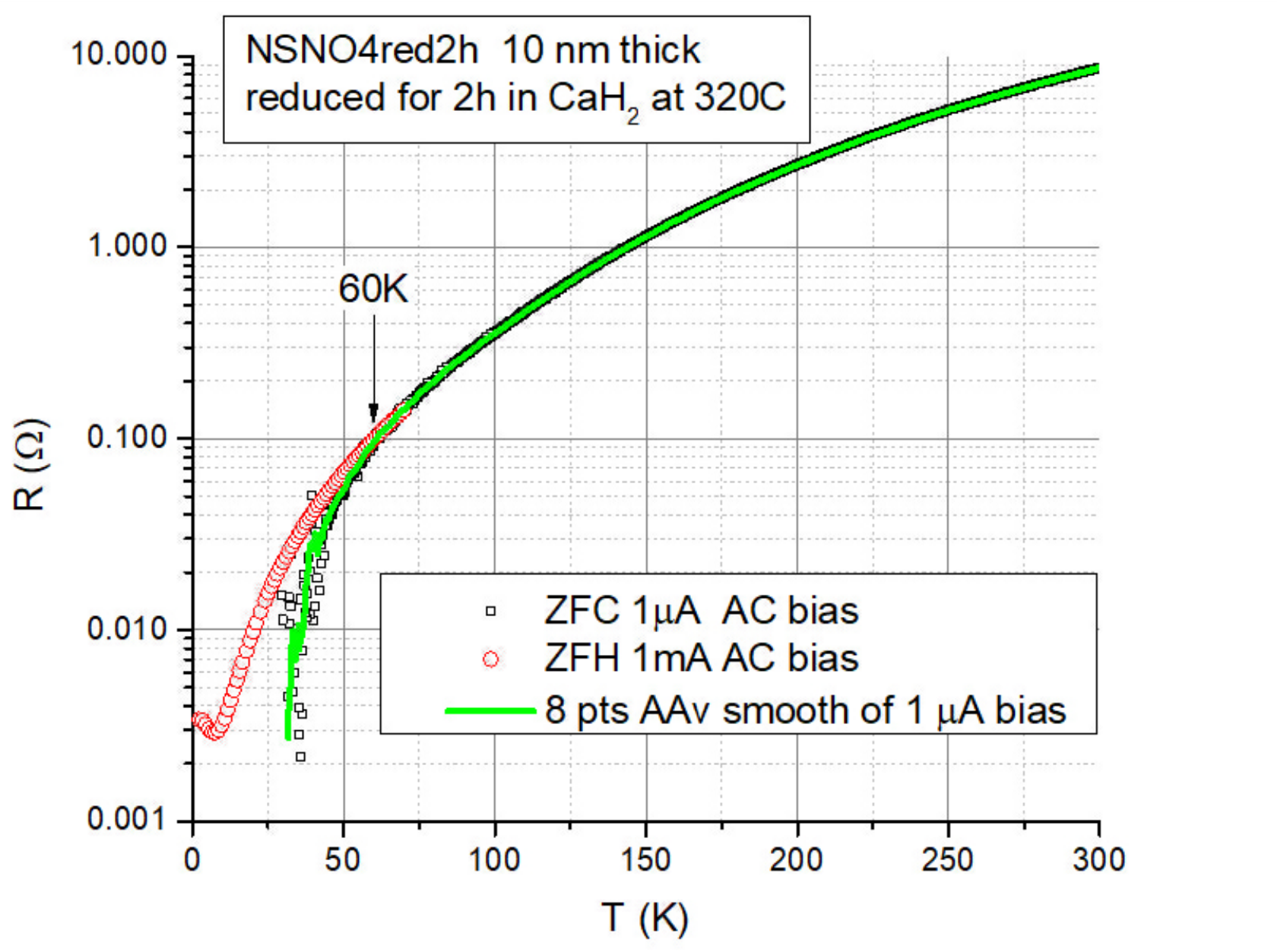}
 \caption{R versus T of NSNO4red2h, a 10 nm thick film, as measured by the PPMS system at 1 $\mu$A and 1 mA bias. At low temperatures, the IVCs are linear only very close to zero bias due to the diode effect (see Figs. 7 and 8 of the main article). Therefore, the 1 mA bias result at low temperatures is misleading and shows finite resistance below 30 K. Thus, the noisy data under 1 $\mu$A bias at low temperatures is more reliable. It shows a transition onset (deviation from the 1 mA bias result) already at about 60 K.   }
\label{figS4}
\end{figure}

In addition to the 30 and 80 nm thick films described in the main article, we also prepared films of 10 and 20 nm thickness. These films were not as extensively characterized, but here in Fig. S3 we present magnetization moment measurements versus temperature of a 20 nm thick film that was reduced for 1.5 h in $\rm CaH_2$ at $320\,^\circ\mathrm{C}$. One can see that the high-temperature transition at $\sim 50$ K is well reproduced, supporting the notion of a robust superconducting transition at 50 K. The inset to this figure shows the net magnitude of the moments between 20 and 50 K of this film versus reduction time, as well as those of the main article. One can see that these net moments are almost linear in the duration of their reduction time and almost independent of the films' thicknesses in this range. This result indicates that it is most likely that the moments originate at the interface between these films and their substrates, regardless of their thickness down to 20 nm. Figure 4 of the next section 4 shows that this is true also down to 10 nm. 

\subsection{ R vs T of a 10 nm thick film}

Though we do not have magnetic moment results on our 10 nm thick film (NSNO4red2h, reduced in $\rm CaH_2$ at $320\,^\circ\mathrm{C}$ for 2 h), we do have transport results on it. Figure S4 depicts two resistance versus temperature results under low and high bias currents as measured by the PPMS system. We measured many IVCs using this system on this film and found very nonlinear behaviors at low temperatures. Nevertheless, at very low bias the IVCs were almost linear, but quite noisy. Therefore, the noisy $R$ vs T here at low T and 1 $\mu$A bias, which we ignored at the beginning, is actually the more reliable result. If we consider the 1 mA bias result, down to about 50 K, as a background mostly due to the black STO substrate, then any deviation from it marks the onset of a possible superconducting transition. One can see that such a deviation of the 1 $\mu$A biased curve occurs at about 60 K. Without the data of the main article, we would have continue to ignore the results of Fig. S4 here, but this data actually further support the existence of a superconducting transition above 50 K (60 K here), as was clearly demonstrated in the main article.

\subsection{Magnetic background effects }

The present state understanding of the phase diagram of nickelates is that their superconductivity coexists with a background of various magnetic orders such as spin glass, anti-ferromagnetism and ferromagnetism \cite{Saykin-Kapitulnik,HaiLin,Shi,Fowlie}. This coexistence occurs to a low extent in low-doped cuprates, is more prominent in the iron pnictides, and even more so in the nickelates. In the cuprates, this occurs as a result of antiferromagnetic ordering of the Cu$^{2+}$ spins in the $\rm CuO_2$ cuprate planes, while in the pnictdes and nickelates the Fe and Ni ions are intrinsically magnetic in nature. The question is how these magnetic orders affect superconductivity in these systems. In the cuprates it was found that magnetic ions such as Gd or Eu when included the 123 phase crystal structure, have no adverse effects on superconductivity. This seemingly surprising result originates from their electronic configuration in the crystal that renders them non-magnetic in the 123-crystal phase. Phase separation that produces magnetic puddles in the material can, of course, affect superconductivity by lowering its $\rm T_c$, broadening the transition, or rendering the material non-superconducting at all. This scenario is quite likely to occur in the nickelates where crystal disorder and competing phases are quite prevalent. \\

\renewcommand{\thefigure}{S\arabic{figure}}

\begin{figure}
 \centering
        \includegraphics[width=1\textwidth]{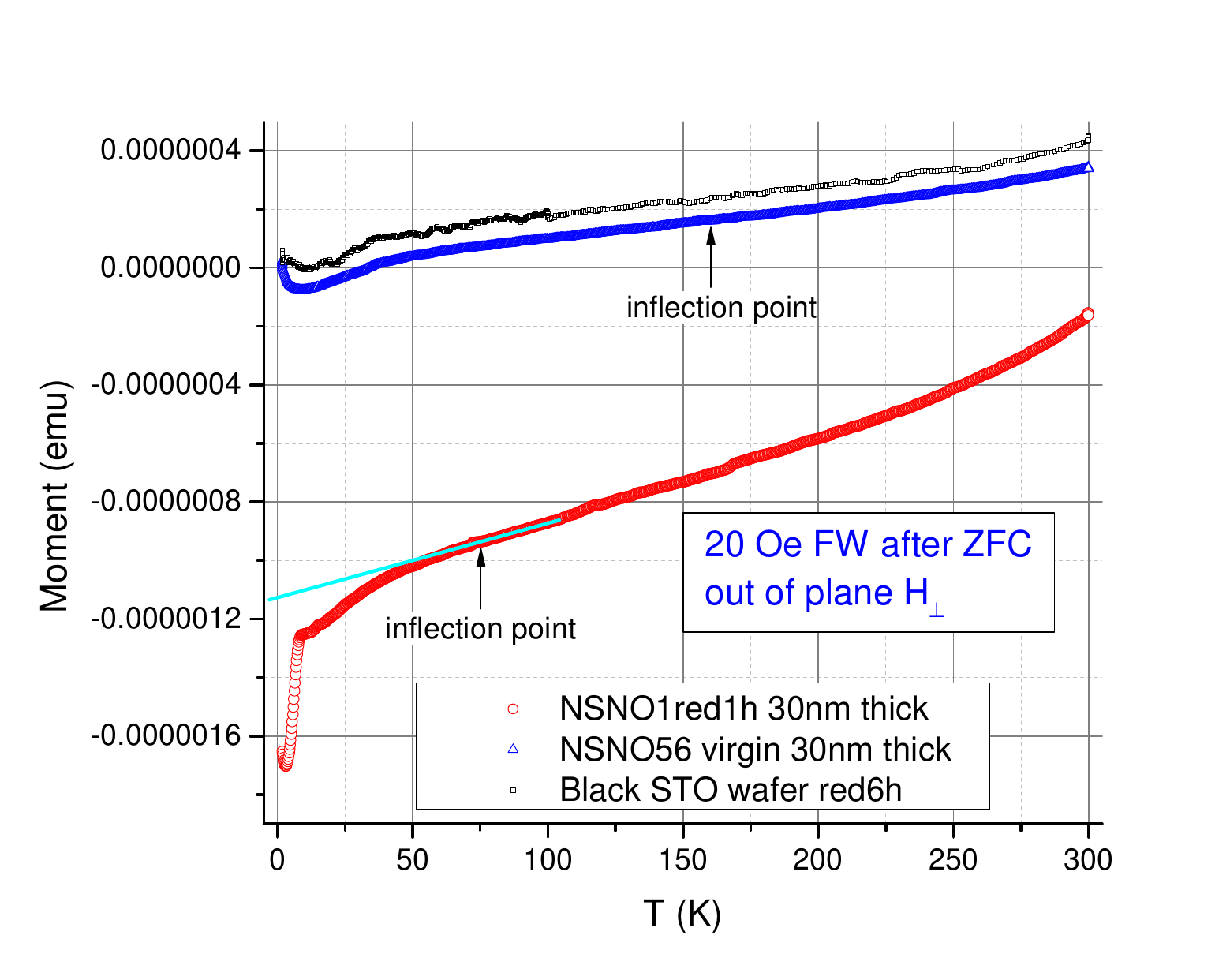}
 \caption{ The magnetic moment M under 20 Oe versus T  up to 300 K, of 30 nm thick NSNO1red1h (same film as in Fig. 2 (b)), a new 30 nm thick virgin film NSNO56, and of a black STO wafer STOred6h (background).   }
\label{figS5}
\end{figure}

As discussed in the main article in connection with Fig. 2 (b), we already showed that the low $\rm T_c$ transition to superconductivity is well established, and that the higher temperature one at about 50 K is clearly diamagnetic and most likely originates in superconductivity. Nevertheless, in view of the magnetic background of the nickelates as discussed above, we discuss specifically how it might affect our results. To do so, we present in Fig. S5 a comparison of M vs T results of three of our samples up to 300 K. One is the 30 nm thick infinite-layer film NSNO1red1h of Fig. 2 (b) of the main article, the second is that of a virgin as deposited 30 nm thick perovskite NSNO film, and the third is that of a black STO wafer which serves as a background to the former two films. The origin of the broad transition of NSNO1red1h at 50-70 K which is close to the inflection point of M vs T at 75 K, might be in the short-range antiferromagnetic order as suggested by Fowlie \textit{et al.} \cite{Fowlie}. The observed transition however is diamagnetic in nature, thus ruling out that the AF order if any, affects it by much. As explained before in the main article, this diamagnetism also behaves oppositely to the observed M vs T of a spin-glass in which M decreases with increasing temperature above the low $\rm T_c$ transition when the Meissner protocol (FW after ZFC) was followed \cite{Saykin-Kapitulnik,HaiLin}. We thus conclude that the background magnetic orders in our films do not affect our results in a significant way. \\

Figure S5 also shows a big difference between the absolute (negative) magnetization of the reduced and virgin films. At 10 K, the magnetization of the virgin film is about 5\% of that of the reduced film. Therefore, its contribution to M of the reduced film, if any, is negligible. It's M vs T is also quite similar to that of the bare STO wafer, and both have a small transition-like behavior at 35-50 K, reminiscent of the transition at 50-70 K in the reduced film. Possibly, there is a precursor to this transition already in the virgin perovskite film. At higher temperatures, the virgin film shows an inflection point at around 150 K. This might indicate an antiferromagnetic transition as reported in the literature \cite{Catalano18} and also observed in muon spin rotation experiments \cite{Fowlie}. The inflection point in the reduced film occurs at a much lower temperature of 75 K, and below it M drops much faster with decreasing temperature compared to the virgin film. This might indicate a different origin of the transition, not to AF order but most likely to superconductivity. Moreover, the strong decrease in M below the transition onset at 50-70 K seems to support Meissner diamagnetism of a superconductor.\\

In conclusion, the magnetic background orders in the nickelates seem to coexist with the $\rm T_c$ enhancement to 50-70 K observed here in our reduced films on  highly reduced STO substrates.  That this transition at 50-70 K actually originated in superconductivity, was demonstrated by its diamagnetic Meissner response, by its significant resistance drop below its onset, and by its link to the observed paramagnetic Meissner peaks just below it, all in the same temperature range.

\subsection{Complementary STEM and EDS results of our films}

This section presents STEM and EDS results of a fresh virgin $\rm Nd_{0.8}Sr_{0.2}NiO_3$ film NSNO57 (measured by a 200 kV Spectra microscope), as well as results measure on the 3-4 months aged NSNO2red1.5 and NSNO2red6h films of Figs. 1 and 4 (measured by a 300 kV Titan microscope).\\

The data on the virgin film is needed for comparison with the corresponding data of the reduced films presented in Figs. 1 and 4. We wished to answer two questions. One concerning the Ti diffusion into the films. Is this effect a result of the reduction process? or was it already present in the parent perovskite film? The second question concerns epitaxy, disorder and defects in the film and to what extent the reduction process affected them. \\

\renewcommand{\thefigure}{S\arabic{figure}}

\begin{figure}
 \centering
        \includegraphics[width=1\textwidth]{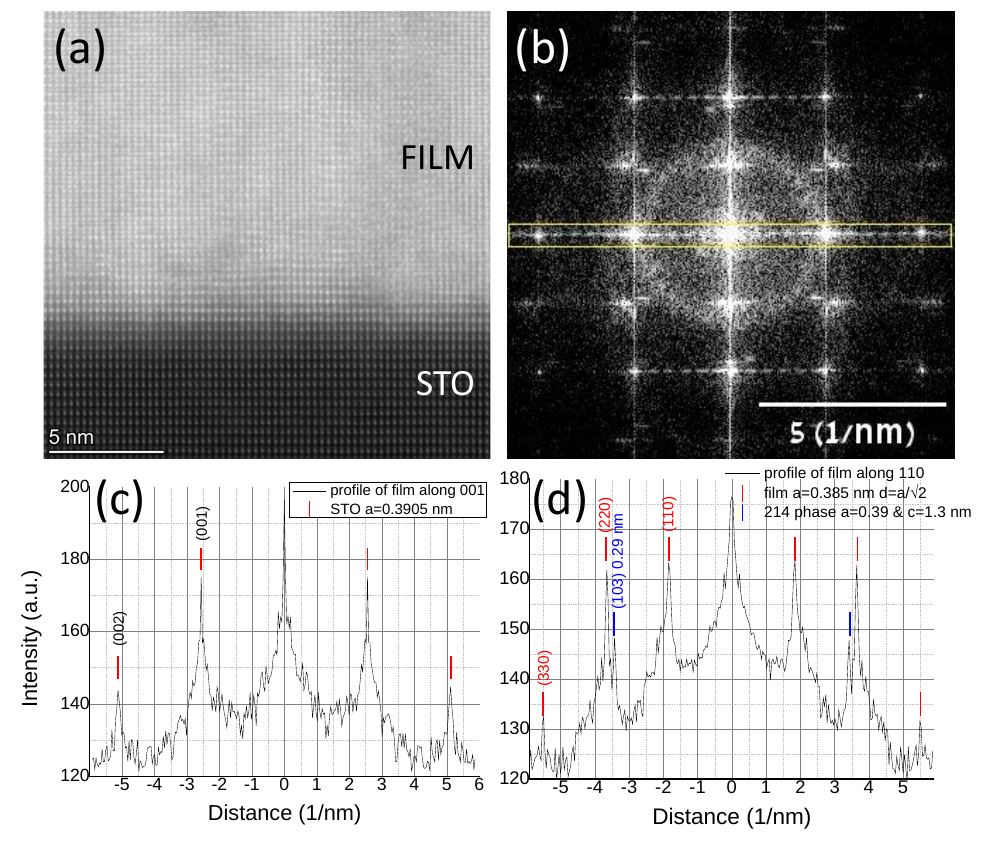}
 \caption{A STEM image along [110] zone axis of a 40 nm thick virgin film of  $\rm Nd_{0.8}Sr_{0.2}NiO_3$ (NSNO57) is shown in (a), while in (b) the FFT of its film only is depicted. In (c) and (d) FFT profiles along [001] and [110] directions, respectively are plotted showing that the dominant phase here is the 113 perovskite. While in (c) we couldn't observe deviations from the peak locations of STO, in (d) a lattice constant of a=0.385 nm compared to 0.381 nm of the bulk was found. This was induced by the tensile strains with the STO wafer that has a=0.3905 nm lattice constant.  }
\label{figS6}
\end{figure}

\renewcommand{\thefigure}{S\arabic{figure}}

\begin{figure}
 \centering
        \includegraphics[width=1\textwidth]{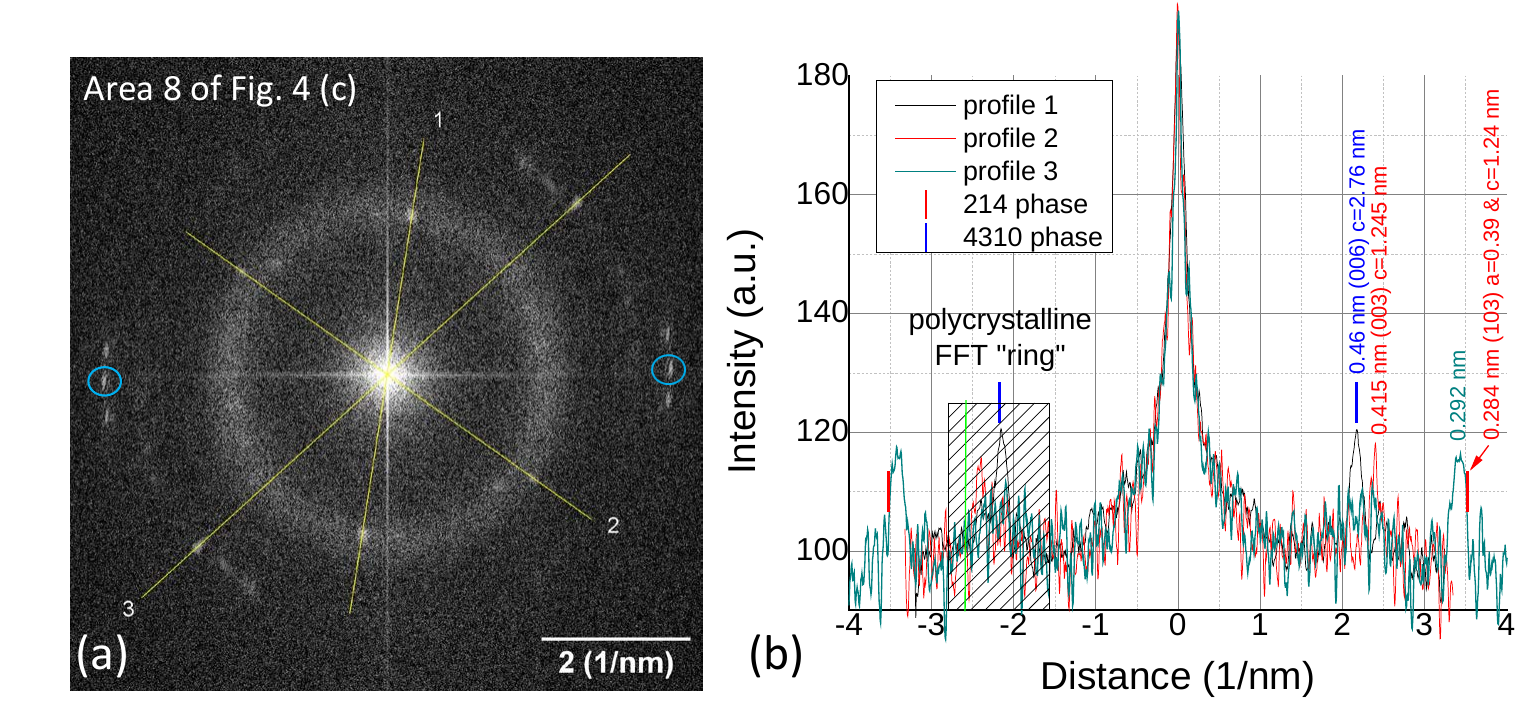}
 \caption{ An FFT of area 8 of the NSNO2red6h film of Fig. 4 (c) is shown in (a) with three lines via stronger spots. Lines 1 and 2 along inner and outer spots on the polycrystalline ring, respectively, and line 3 via farther spots. In (b) the corresponding FFT intensity profiles are plotted. Two of the peaks were found to originate in the 214 phase and one in the 4310 phase. The broad peak due to the polycrystalline ring in (a) is shaded in (b) and includes the 113 RP phase (at the vertical green line) where n is very large. No 113 peak is visible in the profiles in (b), in agreement with the XRD results on our reduced films. The spots circled in blue in (a) at $\pm$3.8 1/nm (0.263 nm) belong to the (008) peak of the 327 phase.      }
\label{figS7}
\end{figure}

Fig. S6 (a) shows a STEM image along the [110] zone axis of a freshly prepared virgin perovskite film under 1.5 J/cm$^2$ laser  fluence. Similar to Fig. 1 (d) and (e), one can see 4-6 ordered epitaxial perovskite layers above the interface, with the vertical columnar defect structures develop further up. FFT of the whole film area is depicted in Fig. S6 (b), where the scale calibration was done using the FFT of the STO with its known lattice constant.  In addition to the strong but fuzzy dots of the mostly epitaxial film, there is also a ring which indicates poly-crystallinity due to the defects and disorder. Intensity profiles along [001] and [110] are plotted in (c) and (d), respectively. They were taken along the narrow yellow rectangle in (b) in order to minimize noise when using line profiles. The [110] profile was obtained by rotating the FFT image by 90$\rm ^o$ and using the same rectangle. \\        

In both profiles in Fig. S6 (c) and (d), the dominant phase is the 113 RP phase with n$\rightarrow \infty$, which comprises basically of many perovskite blocks with occasional Rock salt intergrowth. The peaks in the profile in (c) exactly coincide with those of STO (c=0.3905 nm), while in (d) the in-plane lattice constant that fits the peaks best is a=0.385 nm  with the (110) distance between planes d=a/$\sqrt{2}$. Bulk $\rm NdNiO_3$ has a pseudo-cubic lattice constant of 0.381 nm, thus  a=0.385 nm of the film here indicates that it is under tensile strain created by the larger lattice constant of STO. This in turn should have shrunk the c constant but this wasn't resolved in (c), apparently due to the intergrowth. \\

Figure S7 (a) shows an FFT image of area 8 of the NSNO2red6h film of Fig. 4 (c). It depicts a broad ring indicating its poly-crystallinity, and some higher intensity spots originating in larger ordered crystallite regions, though which three lines are drawn.  FFT intensity profiles along these lines are plotted in (b). The broad shadowed peak in (b) originates in diffraction of the ring, while the three sharper peaks were identified as due to the 214 and 4310 RP phases. The 327 RP phase is also present here, as one can see from the strong spots circled in blue in (a) at $\pm$3.8 1/nm (0.263 nm).  The location of the 113 RP phase (n$\rightarrow \infty$) is marked by the vertical green line in the shadowed region. No sharp peak of this phase is visible at this green line location, indicating the absence of large ordered crystallites of this phase. This absence is also in agreement with our XRD results on this highly reduced film (see Fig. S1). No IL phase was observed in (b) either, plausibly due to aging and oxygen intake by this film.\\

Fig. S8 summarizes the results of some representative FFT and intensity profiles of areas 7, 2 and 3 of Fig. 4 (c) and (d), and also of the whole film in (d). (a) shows that in the dark area 7 there are large regions with single crystals of the 327 phase, oriented in the same direction. In Fig. S8 (b) of the whole film in Fig. 4 (d), the 214 phase could be identified. (c) and (d) represent areas 2 and 3 where visible two sets of parallel lines of atoms in perpendicular directions are clearly seen in real space. This give rise to FFT diffraction dots along two diagonals normal to these lines. The FFT intensity profiles along these diagonals show evidence for the existence of the 214 phase in areas 2 and 3. To summarize, the long reduction duration of the NSNO2red6h film, made it polycrystalline with RP phases that we couldn't resolve (in the ring), together with regions of larger ordered crystallites (the more intense dots) which were identified as belonging to the  RP phases with n=1, 2 and 3. \\ 

\renewcommand{\thefigure}{S\arabic{figure}}

\begin{figure}
 \centering
        \includegraphics[width=1\textwidth]{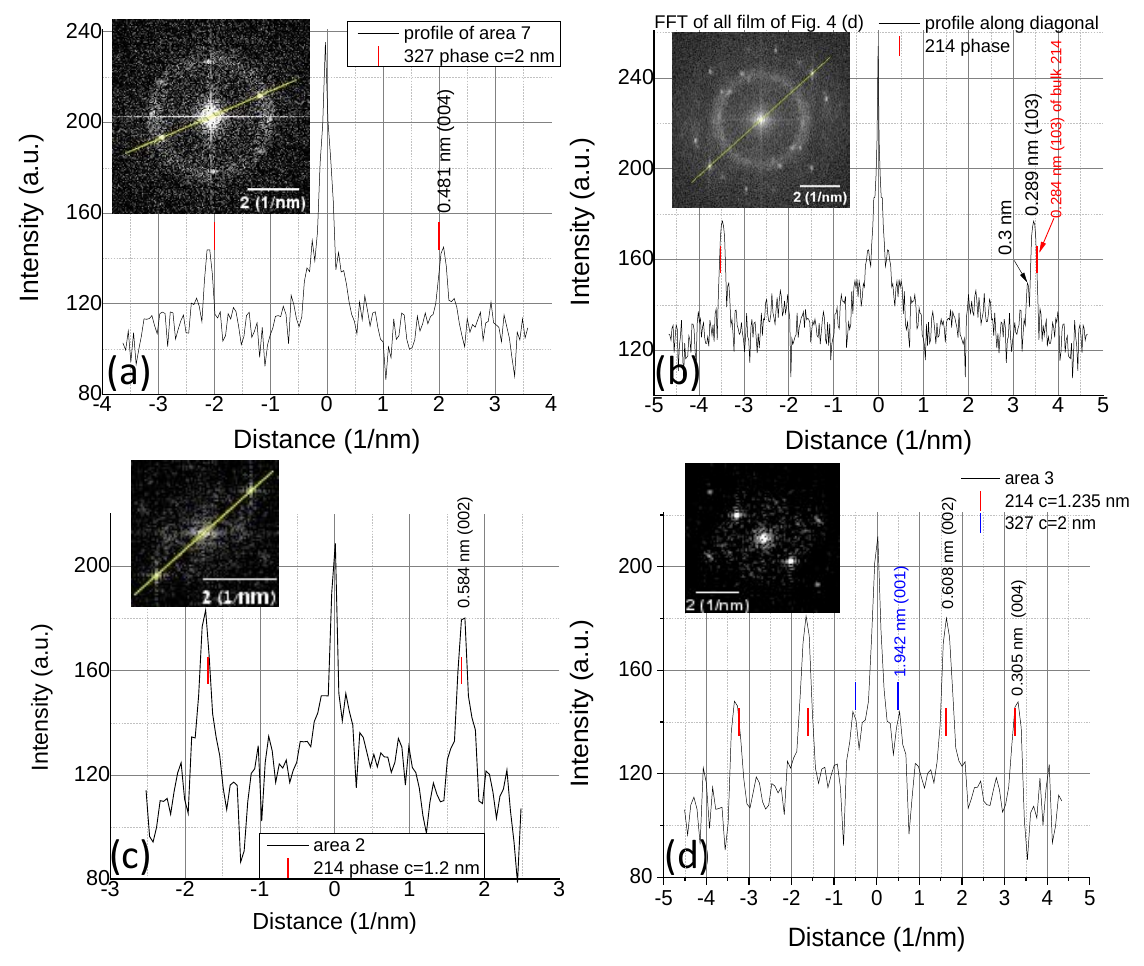}
 \caption{FFT profiles of different areas as marked in Figs. 4 (c) and (d), along the lines shown except for (d) where the profile is along the [-110] orientation (not marked to avoid masking the weak outer spots). The stronger spots here originate in more ordered crystallites of which the 214  and 327  RP phases were identified.    }
\label{figS8}
\end{figure}

Finally, we go back to STEM results measured on the virgin NSNO57 film.  In Figs. S9 and S10 we present EDS results of this film. In Fig. S9, maps of atomic concentration are depicted for the different atoms in the film and substrate, on the area shown by the STEM image on the left. One can see alternating vertical concentration columns in the Nd and Ni maps. This is also observed in the mixed color map (second from left). This phenomenon is related to the vertical columnar disorder seen in Figs. 1 (d) and (e), and S6 (a). Interestingly, in the atomic concentration maps of Fig. 1 (h) of the main article, this columnar structure is smeared, the Ni aggregates in puddles, and this aggregation is even more enhanced in Fig. 4 (g) and (h) where a longer reduction time was used. \\

In Fig. S10, the atomic fraction profiles show that there is already a small Ti concentration in the virgin film, though much smaller than that in the reduced films of Figs. 1 (g) and 4 (e). This Ti signal is also very close to the noise level now. We determine the noise level in Fig. S10 as before in Fig. 1 (g), as the mean value of the noise below and above the film, and find that the net Ni+Ti atomic fraction is still equal to that of the  Nd+Sr, to within the noise of the measurement. Thus the stoichiometry of the parent 113 perovskite phase is preserved in the film.\\

\renewcommand{\thefigure}{S\arabic{figure}}

\begin{figure}
 \centering
        \includegraphics[width=1\textwidth]{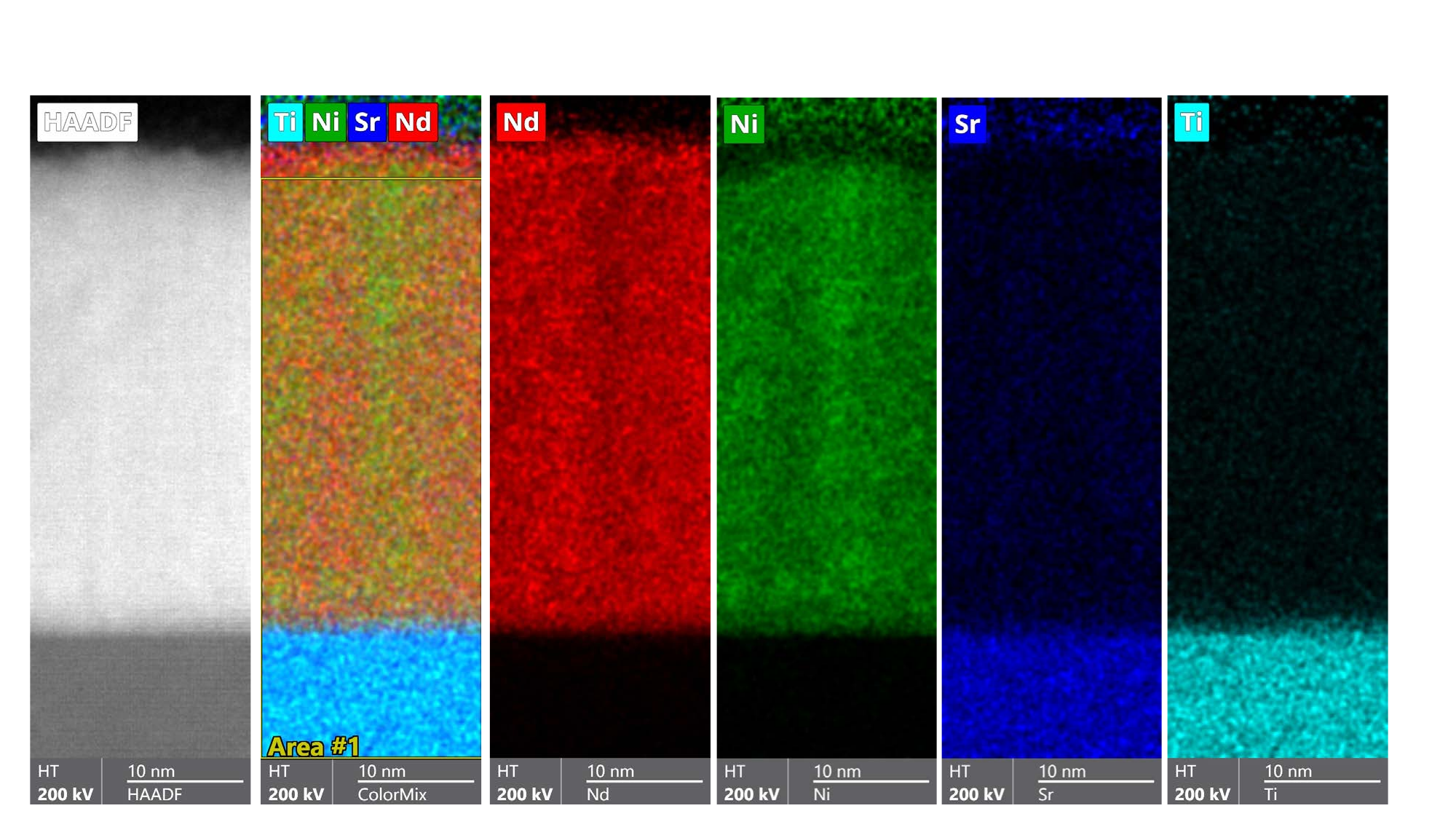}
 \caption{ EDS maps of atomic concentration of the different atoms in our virgin film NSNO57, in the area shown by the STEM image on the left side. In the mixed color map (second from left), and also in the Nd and Ni maps, alternating vertical Nd and Ni concentration columns can be seen similar to the columns structure in Figs.  1 (d) and (e), and S6 (a).  }
\label{figS9}
\end{figure}

\renewcommand{\thefigure}{S\arabic{figure}}

\begin{figure}
 \centering
        \includegraphics[width=1\textwidth]{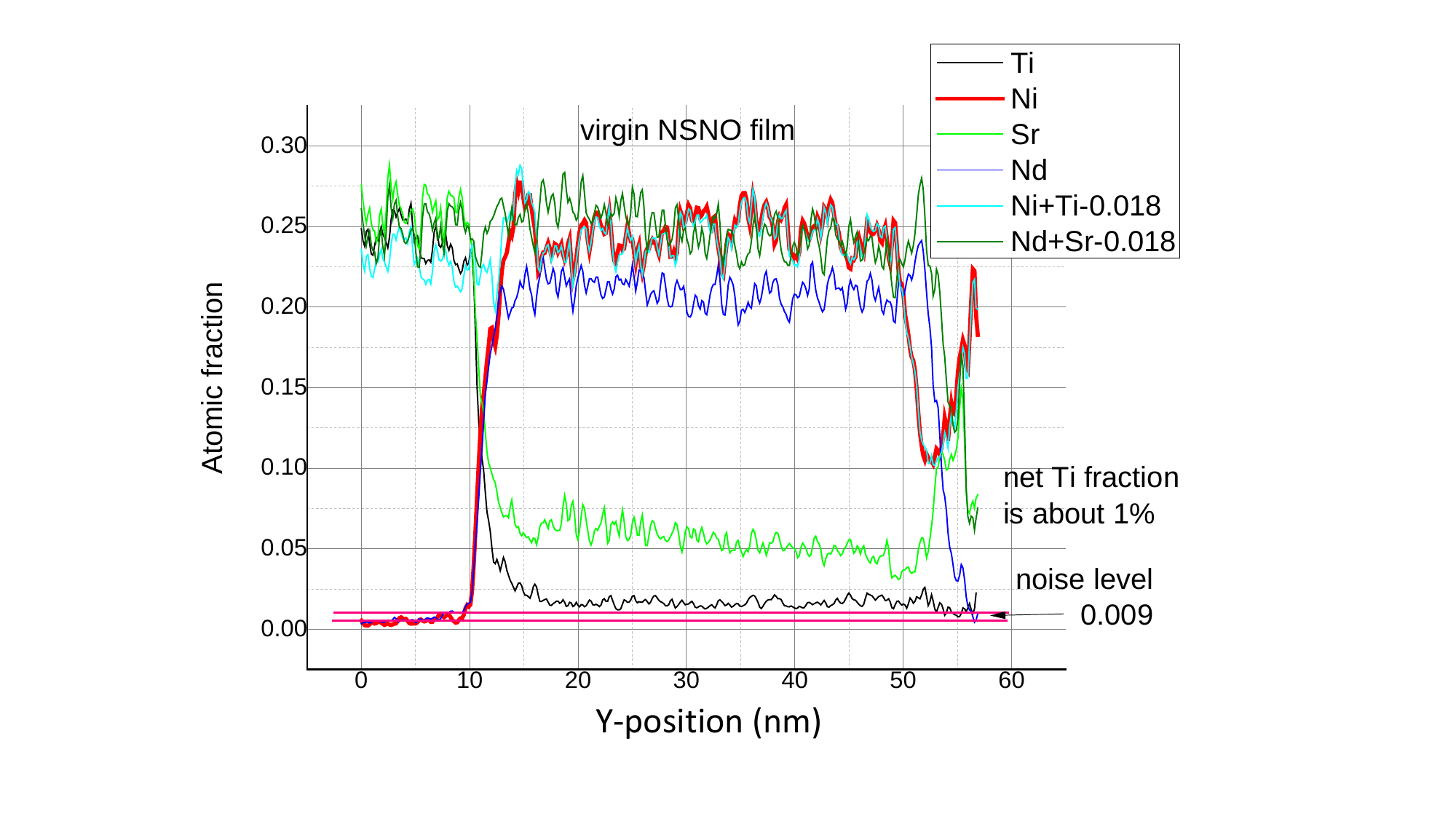}
 \caption{ Atomic fraction profiles of integrated EDS signals over layers parallel to the interface versus the y-position as seen in Fig. S9. The upturned EDS signals of the Ni and Sr above the film are unreliable, since all the net intensity profiles go to zero above the film.   }
\label{figS10}
\end{figure}

\newpage

%\subsection{References}
%\nocite{*}

%\bibliography{NSNO-bibliography}% Produces the bibliography via BibTeX.

\bibliographystyle{iopart-num} %<--- CHANGE THIS FROM apsrev4-1
\bibliography{main.bib}   %<--- Name of your .bib file

\end{document}